\documentclass{aa}

\usepackage{txfonts}
\usepackage{supertabular}
\usepackage{dsfont}
\usepackage{graphicx}
\usepackage{epstopdf}
\usepackage{color}
\usepackage{tabularx}
\usepackage{calc}
\usepackage{array}
\usepackage[authoryear]{natbib}

\begin{document}

\titlerunning{Influence of Planck foreground masks in the large angular scale quadrant CMB asymmetry}

\title{Influence of Planck foreground masks in the large angular scale quadrant CMB asymmetry}

\author{L. Santos \inst{1},  P. Cabella \inst{2,3}, T. Villela \inst{4}, W. Zhao\inst{1,5}}

\institute{CAS Key Laboratory for Researches in Galaxies and Cosmology, University of Sciences and Technology of China, Chinese Academy of Sciences, Hefei, Anhui 230026, China \and Universit\`a di Roma ``Tor Vergata'', Dipartimento di Fisica, Rome, Italy \and INFN, Sezione di Roma Tor Vergata, Rome, Italy \and Instituto Nacional de Pesquisas Espaciais - INPE, Divis\~ao de Astrof\'isica, S\~ao Jos\'e dos Campos, SP, Brazil \and State Key Laboratory of Theoretical Physics, Institute of Theoretical Physics, Chinese Academy of Sciences, Beijing 100190, China}

\offprints{L. Santos \email{larissa@ustc.edu.cn}}

\date{Received      /Accepted}

\abstract {The measured cosmic microwave background (CMB) angular distribution shows a great consistency with the $\Lambda$CDM model, that predicts cosmological isotropy as one of its fundamental characteristics. However, isotropy violations were reported in CMB temperature maps of the Wilkinson Microwave Anisotropy Probe (WMAP) and confirmed by the Planck satellite data.}  
{Our purpose is to investigate the influence of different sky cuts (masks) employed in the analysis of CMB angular distribution, in particular in the excess of power in the Southeastern quadrant (SEQ) and the lack of power in the Northeastern quadrant (NEQ), found in both WMAP and Planck data.}  
{We compare the two-point correlation function (TPCF) computed for each quadrant of the CMB foreground-cleaned temperature maps to 1000 Monte Carlo (MC) simulations generated assuming the $\Lambda$CDM best-fit power spectrum using four different masks, from the less to the most severe one: mask-rulerminimal, UT78, U73 and U66.  In addition to the quadrants and  for a better understanding of these anomalies, we computed the TPCF using the mask-rulerminimal for circular regions in the map where the excess and lack of power are present.  We also compare, for completeness, the effect of Galactic cuts (+/- 10, 20, 25 and 30 degrees above/below the Galactic Plane) in the TPCF calculations as compared to the MC simulations.} 
{We found consistent results for three masks, namely mask-rulerminimal, U73 and U66. The results indicate that the excess of power in the SEQ tends to vanish as the portion of the sky covered by the mask increases and the lack of power in the NEQ remains virtually unchanged.  A different result arises for the new released UT78 Planck mask. When this mask is applied, the NEQ becomes no longer anomalous. On the other hand, the excess of power in the SEQ becomes the most significant one among the masks. Nevertheless, the asymmetry between the SEQ and NEQ is independent of the mask and it is in disagreement with the isotropic model with at least 95\% C.L.} 
{We find that UT78 is in disagreement with the other analyzed masks, specially considering the SEQ and the NEQ individual analysis. Most importantly, the use of UT78 washes out the anomaly in the NEQ. Furthermore, we found excess of kurtosis, compared with simulations, in the NEQ for the regions not masked by UT78 but masked by the other masks, indicating that the previous result could be due to non-removed residual foregrounds by UT78.}


\maketitle

\section{Introduction}

The cosmic microwave background (CMB) radiation is one of the best cosmological observables and provides a powerful test to the so-called standard cosmological model, also known as the $\Lambda$CDM model. The Planck satellite is the fourth generation space mission devoted to CMB measurements and has recently released the most accurate CMB full-sky dataset to date. These results show an outstanding consistency with the spatially flat six-parameter standard model \citep{2015plancka}. 

Homogeneity and isotropy are fundamental properties of the $\Lambda$CDM cosmology, however deviations from statistical isotropy have been reported in CMB data throughout the years.  One of these anomalies was first reported in the Cosmic Background Explorer (COBE) data \citep{1992smoot} and later confirmed by the Wilkinson Microwave Anisotropy Probe (WMAP) observations \citep{2003bennett.1, 2007hinshaw,2009hinshaw,2011jarosik}: a low quadrupole amplitude was detected, in disagreement with the predicted value from the standard model of cosmology.  Other violations of isotropy were soon announced in WMAP data, including an alignment of the low order multipoles  \citep{2004bielewicz,2004schwarz,2004copi,2004deoliveiracosta,2005bielewicz,2005land,2006copi,2006abramo, 2010gruppuso, 2010frommert}, the cold spot \citep{2004vielva,2005cruz,2007cruz,2010vielva.2}, the parity asymmetry \citep{2010kim, 2011gruppuso, 2012aluri, 2012hansen, 2012naselsky, 2014zhao}  and  the North-South asymmetry \citep{2004eriksen,2004hansen,2004eriksen.2,2004hansen.2,2005donoghue,2009hoftuft,2010paci,2010pietrobon,2010vielva}. Closely related to the North-South asymmetry is the power asymmetry found between different quadrants of the CMB sky  \citep{2012santos, 2014santos}. 

Most of the anomalies found in previous CMB observations were confirmed by Planck data suggesting that they are not due to systematic effects in the instruments. Besides the confirmation from Planck team \citep{2014plancka}, \cite{2014bernui} confirmed the North-South asymmetry in Planck data with a 98.1\% C.L. and considered it unlikely to be due to residual foregrounds. Other anomalies were also confirmed by different authors, see, for instance, \cite{2015polastri} and \cite{2014gurzadyan}. However, the significance of CMB anomalies are far from being consensus. On the other hand, results from \cite{2015quartin} show that no significant power asymmetry is present in CMB data when both Doppler and aberration effects are properly removed. \cite{2014rassat} also claimed that, after removing astrophysical and cosmological secondary effects, only the low quadrupole remains anomalous. The authors also concluded that masking the sky to avoid residual foregrounds has a bigger impact on CMB statistics than full-sky CMB analysis. It is essencial to understand the origin of such features in CMB temperature distribution in order to either confirm the $\Lambda$CDM model or search for different explanations from the perspective of a new physics. 

In this paper  we examine how different sky cuts (from more conservative masks to less conservative ones) affect the excess of power found in the Southeastern quadrant (SEQ)  and the significant lack of power found in the Northeastern quadrant (NEQ) of the sky.  We use Planck 2015 foreground-cleaned temperature maps (for a detailed description see \cite{2015planckb}) and different masks as it will be described in Section 2, in addition to the method we use to analyze the data. In section 3 and 4, we discuss the results and present our conclusions.

\section{Method}

Our first step was to generate 1000 Monte Carlo (MC) simulated CMB maps with $N_{side}=256$ using HEALPix (Hierarchical Equal Area and Isolatitude Pixelization) package (Synfast) \citep{2005gorski}.  These simulated CMB maps were created using the best-fit $\Lambda$CMD model power spectrum from Planck \citep{2013planckb}.  As for CMB data, we consider the Planck 2015 release foreground-cleaned temperature sky maps, in particular the SMICA2 (the other maps are NILC2, SEVEM2 and Commander2).  Due to computational limitations and taking into account that we are only concerned with large angular scales, all  maps were degraded to $N_{side}=64$ (including the MC maps).  We then divided every simulated map as well as every CMB map in four quadrants: SEQ, NEQ, Southwestern (SWQ) and Northwestern (NWQ). Furthermore, instead of dividing the sky in quadrants, we also chose circular regions (in CMB map as well as in the simulations) where we found the excess and lack of power in CMB sky.

At this stage, we mask pixels that can still be contaminated by residual foregrounds in every CMB map. Since we are comparing the real data with simulations, the same procedure must be done to the MC maps. Four different masks with different sky cuts are considered. The masks provided by the Planck team are three: from the first data release, mask-rulerminimal  and U73 (they cut 16.35\% and 25.17\% of the sky for $N_{side}=64$, respectively); and from the second data release, mask UT78 (cuts 21.33\% of the sky for $N_{side}=64$). The forth mask is named U66 and is the most severe one (cuts 32.59\% of the sky for $N_{side}=64$). It was constructed by \cite{2014axelsson} and it is publicly available. 

In the case where we considered circular regions, due to computational limitations, we used only the least severe mask (mask-rulerminimal). We also tested different radius for each chosen region. Since we are considering large angular scale asymmetries and for a direct comparison with the previously chosen quadrants (in the number of pixels used in the analysis), we restrict ourselves to radius that run from $60^{\circ}$ to $80^{\circ}$.

The third step is to calculate the two-point correlation function (TPCF) for each quadrant or circular region in every map (CMB data and simulations), applying each of the described masks at a time (in the case where the sky is divided in quadrants). We define the TPCF as the average product between the temperature of all pairs of pixels separated by an angular distance $\gamma$ in each analyzed masked sky map: 

\begin{equation}
c(\gamma)\equiv \langle T({\bf n_p}) T({\bf n_q})\rangle.
\end{equation}

The temperature fluctuations of the pixels $p$ and $q$ are described by $T({\bf n_p})$ and $T({\bf n_q})$, respectively. Moreover, these pixels are defined by the coordinates ($\theta_p$, $\phi_p$) and ($\theta_q$, $\phi_q$), where $0^\circ\leq \phi \leq 360^\circ$ and $-90^\circ \leq \theta \leq 90^\circ$. The angular distance between two generic pixels is given by

\begin{equation}
\cos\gamma = \cos\theta_p \cos\theta_q + \sin\theta_p \sin\theta_q \cos(\phi_p-\phi_q).
\end{equation}

The next step is to quantify the results by calculating a rms-like quantity, $\sigma$,  defined in \cite{2006bernui} for each TPCF curve:
\begin{equation}
\label{sigma}
	\sigma = \sqrt{\frac{1}{N_{bins}}\sum_{i=1}^{N_{bins}}f_i^2},
\end{equation}
\noindent where $f_i$ corresponds to the TPCF  for each bin $i$, being $N_{bins}=90$.

Finally, we compare both the TPCF curves and the rms-like quantity $\sigma$ for Planck CMB maps and for the simulated $\Lambda$CDM model MC maps in each quadrant using four different sky masks. We discuss the results in the forthcoming section.  

\section{Results and discussion}

\begin{figure}
 \includegraphics[scale=0.5]{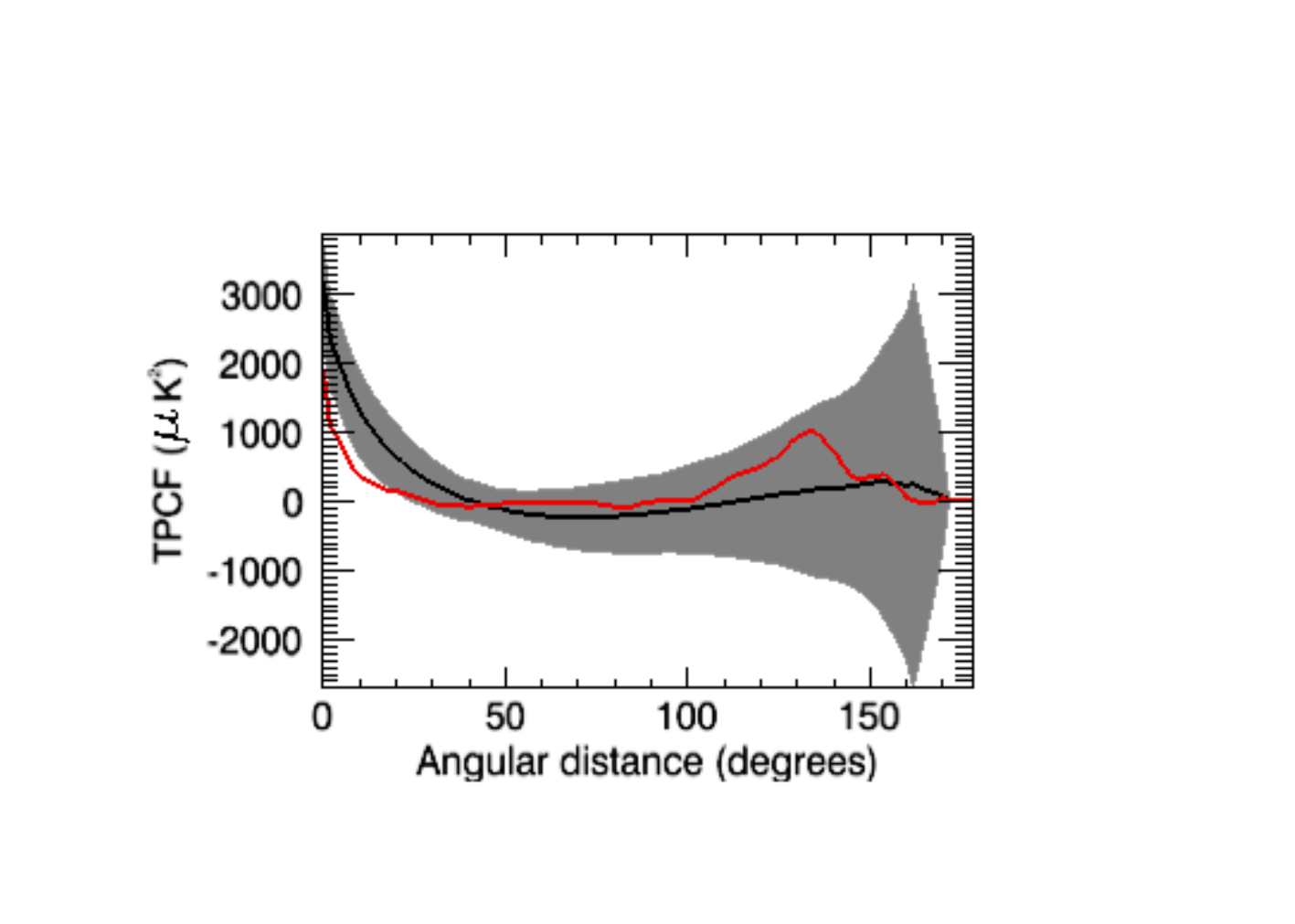}
 \includegraphics[scale=0.5]{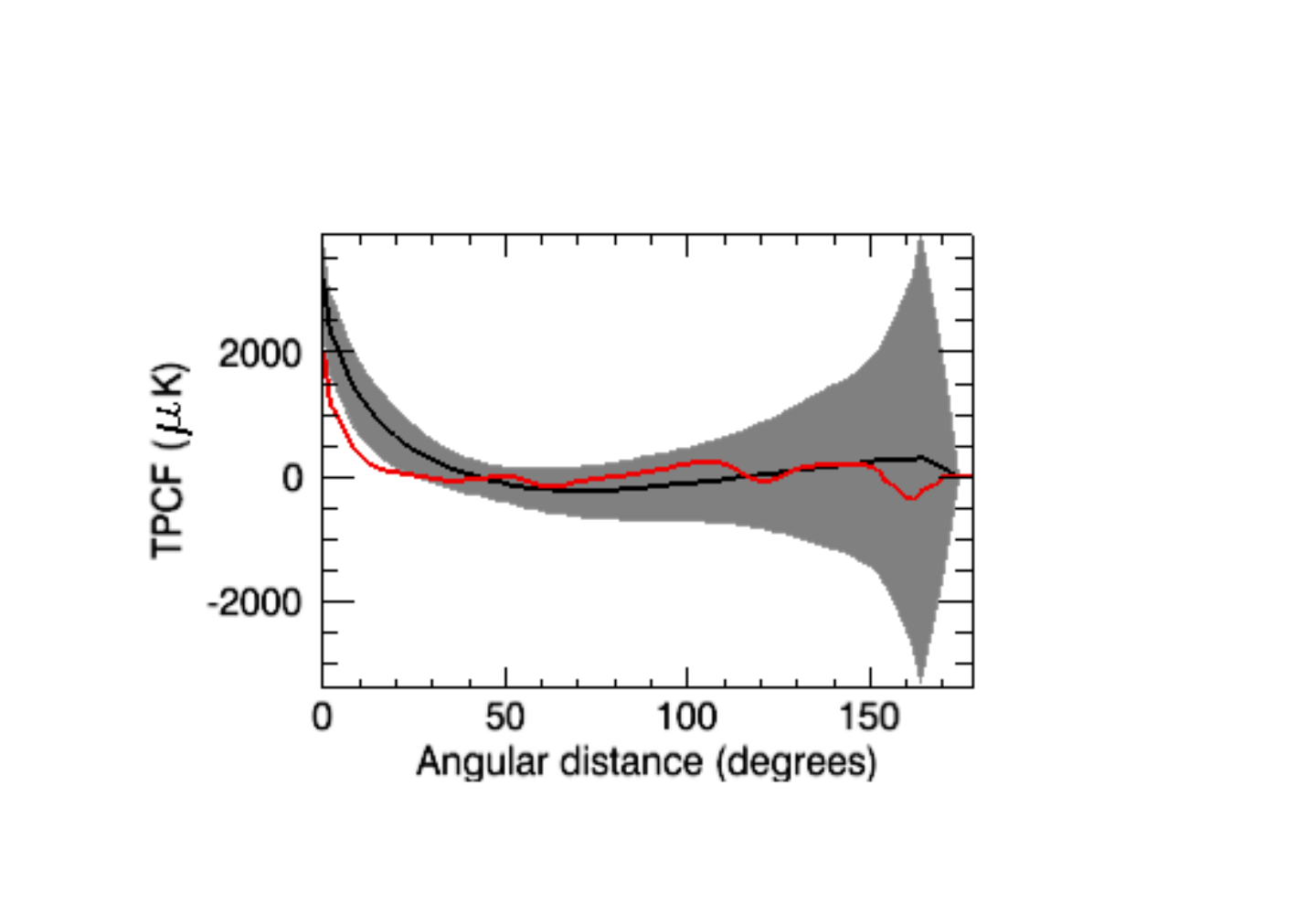}
 \includegraphics[scale=0.5]{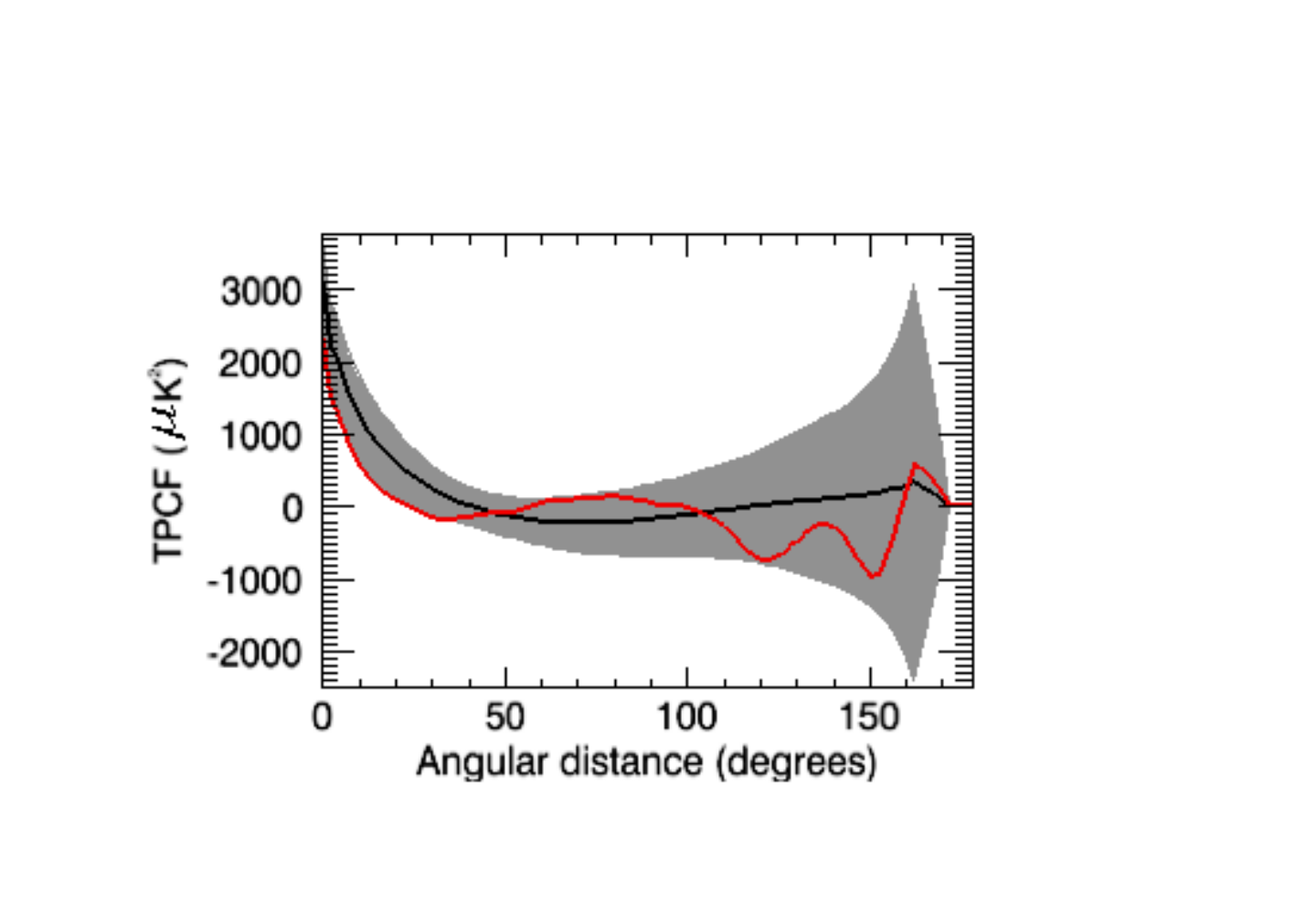}
 \includegraphics[scale=0.5]{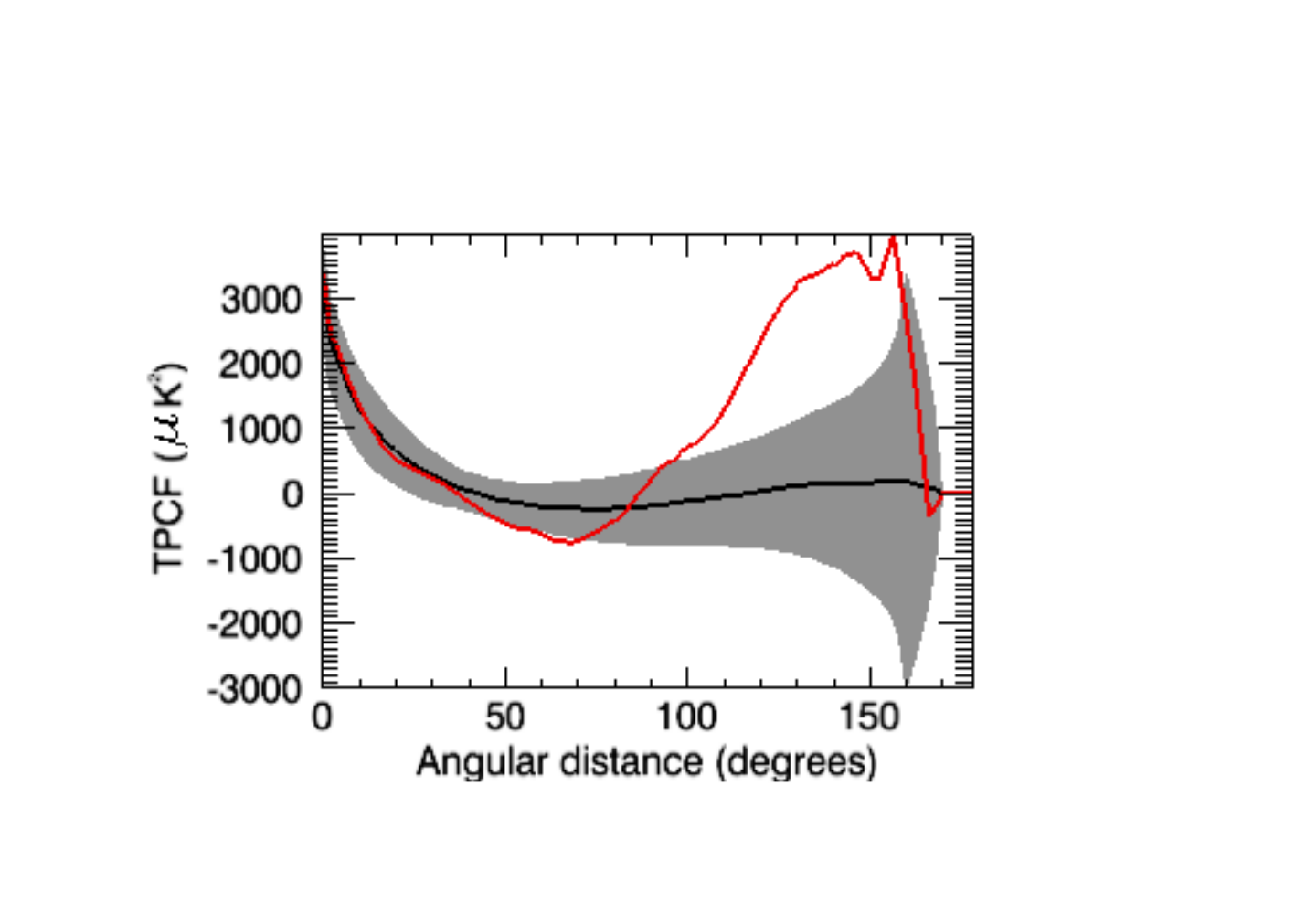}
\caption{TPCF curves computed for the main Planck second data release temperature  foreground-cleaned map (SMICA2) (red curve) using mask-rulerminimal.  We smoothed the curves using the smooth function from Interactive Data Language (IDL) for illustration purposes only (in the calculations we use the original calculated values for the TPCF). From top to bottom, NWQ, NEQ, SWQ, and SEQ appear as solid red lines. The shadow part depicts the standard deviation intervals  (68\% C.L)  for 1000 simulated maps produced with the $\Lambda$CDM spectrum. The black curve is the mean TPCF considering the MC simulated maps.}
\label{TPCF-rulerminimal}
\end{figure}

\begin{figure}
 \includegraphics[scale=0.5]{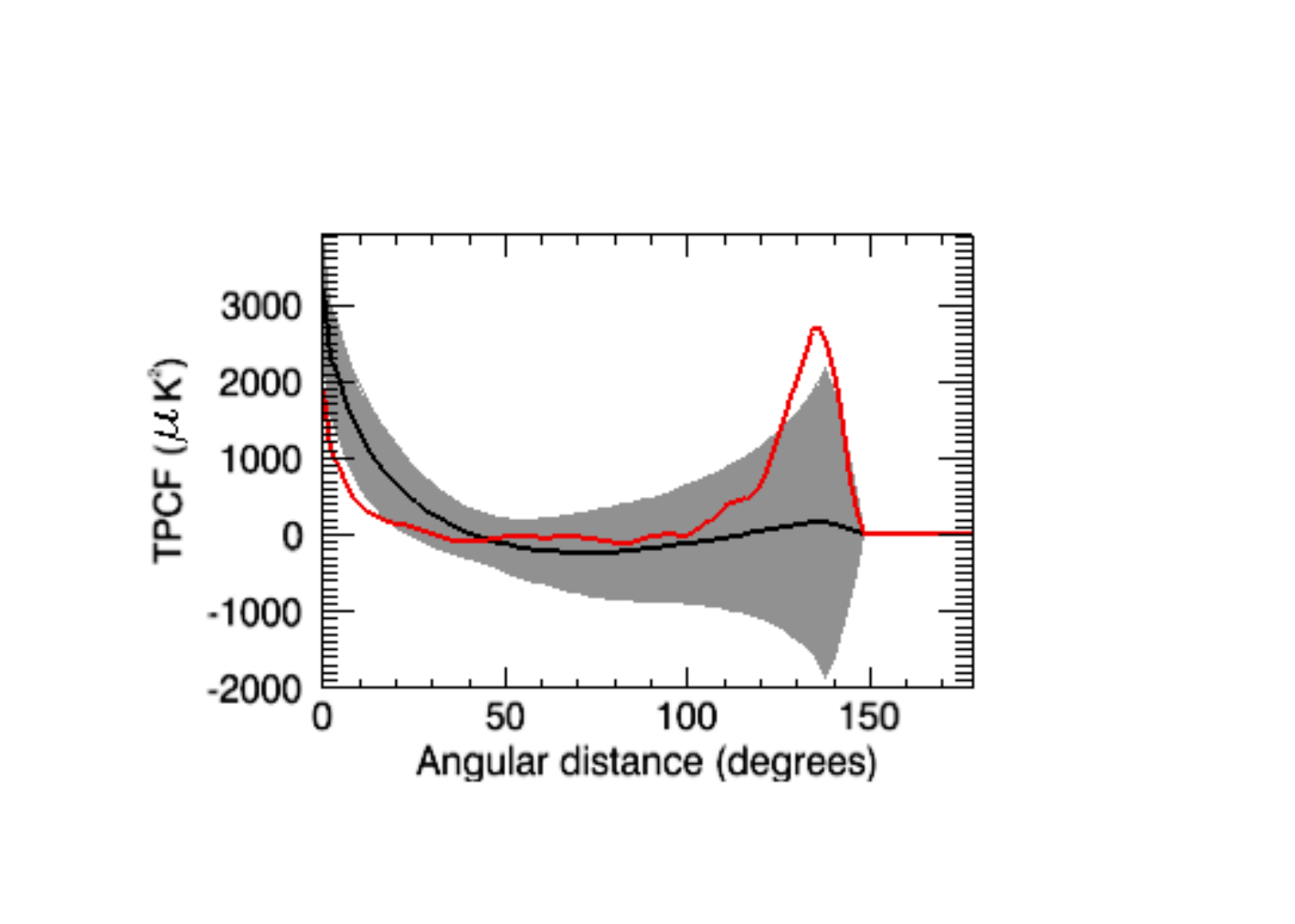}
 \includegraphics[scale=0.5]{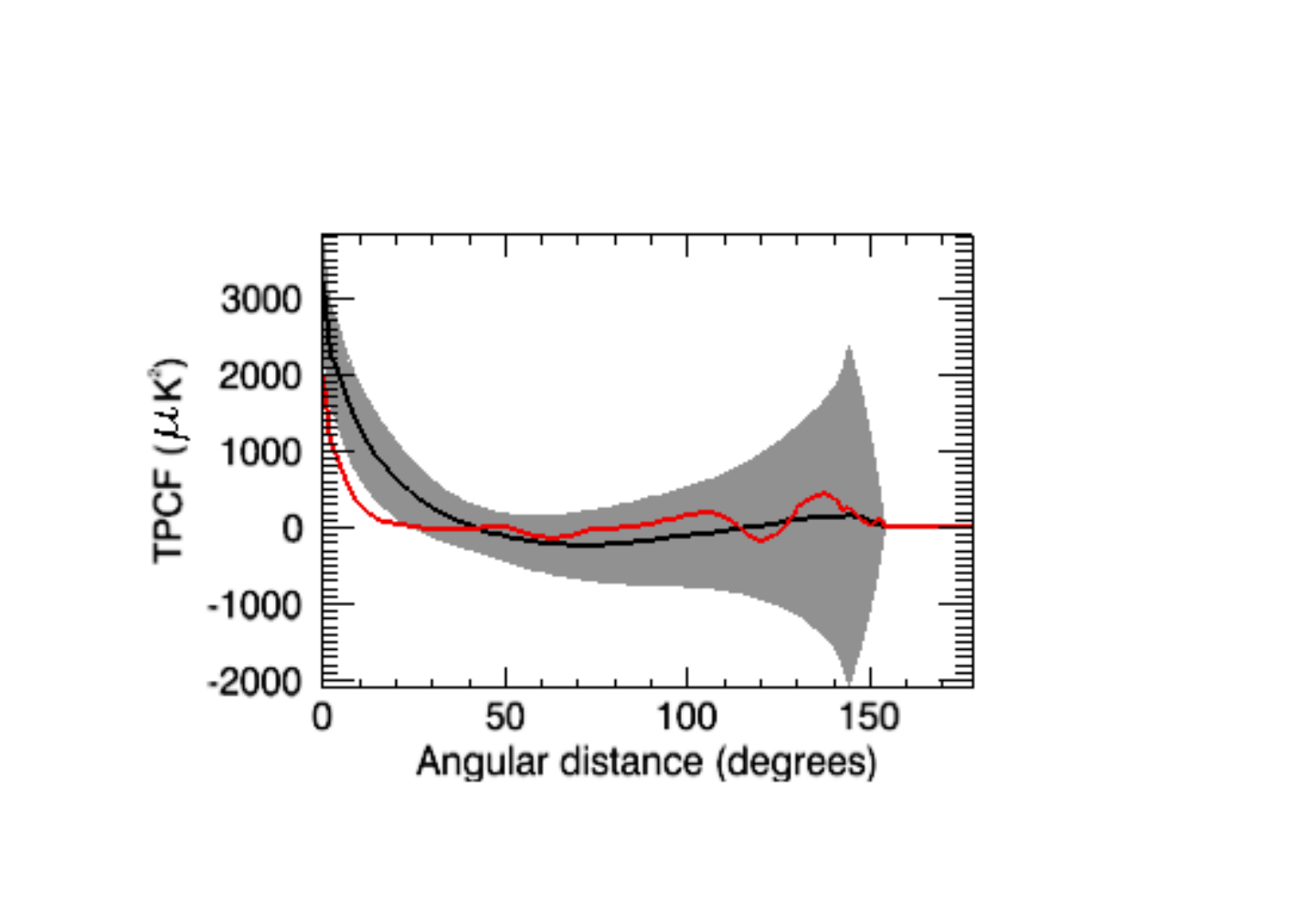}
 \includegraphics[scale=0.5]{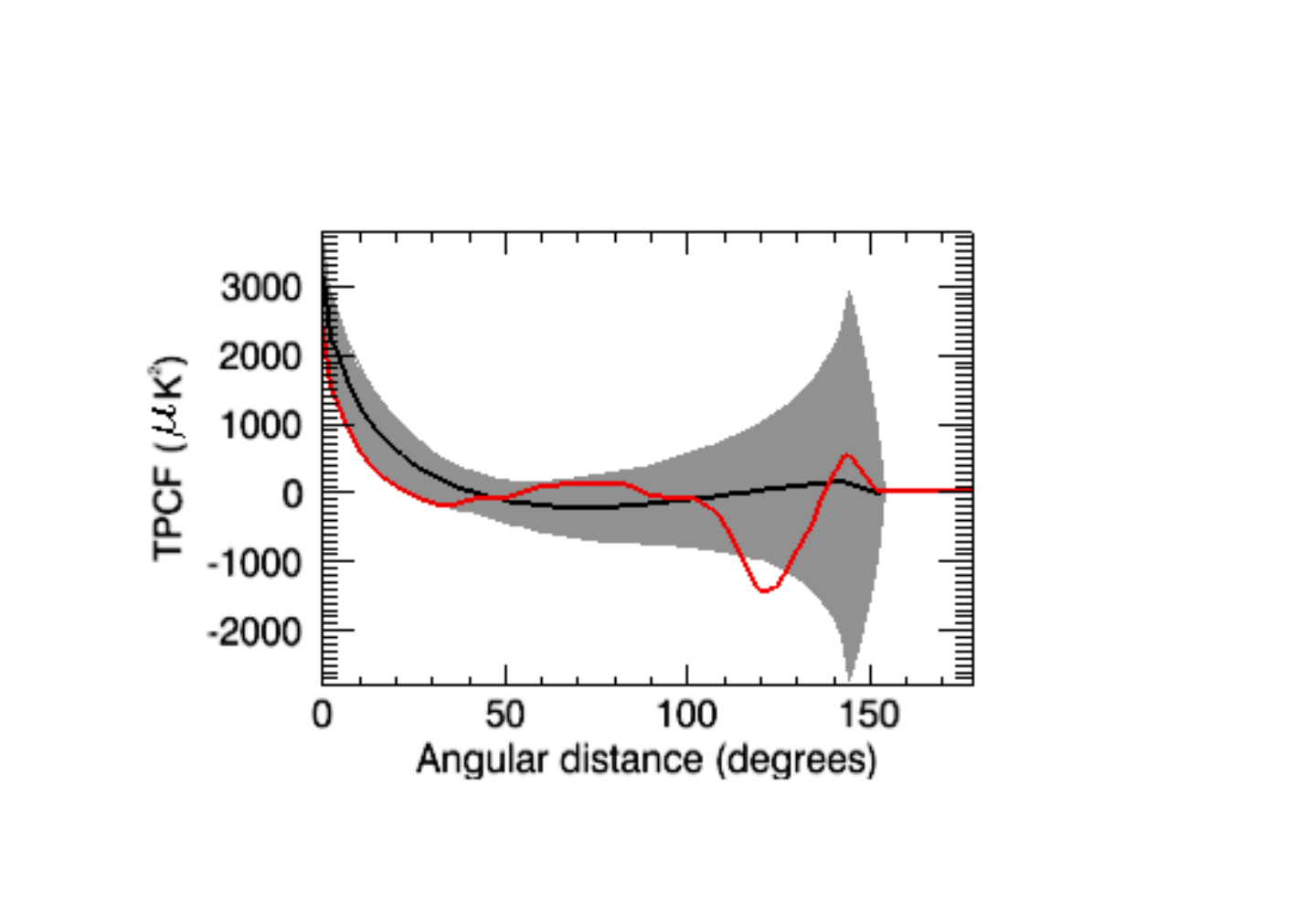}
 \includegraphics[scale=0.5]{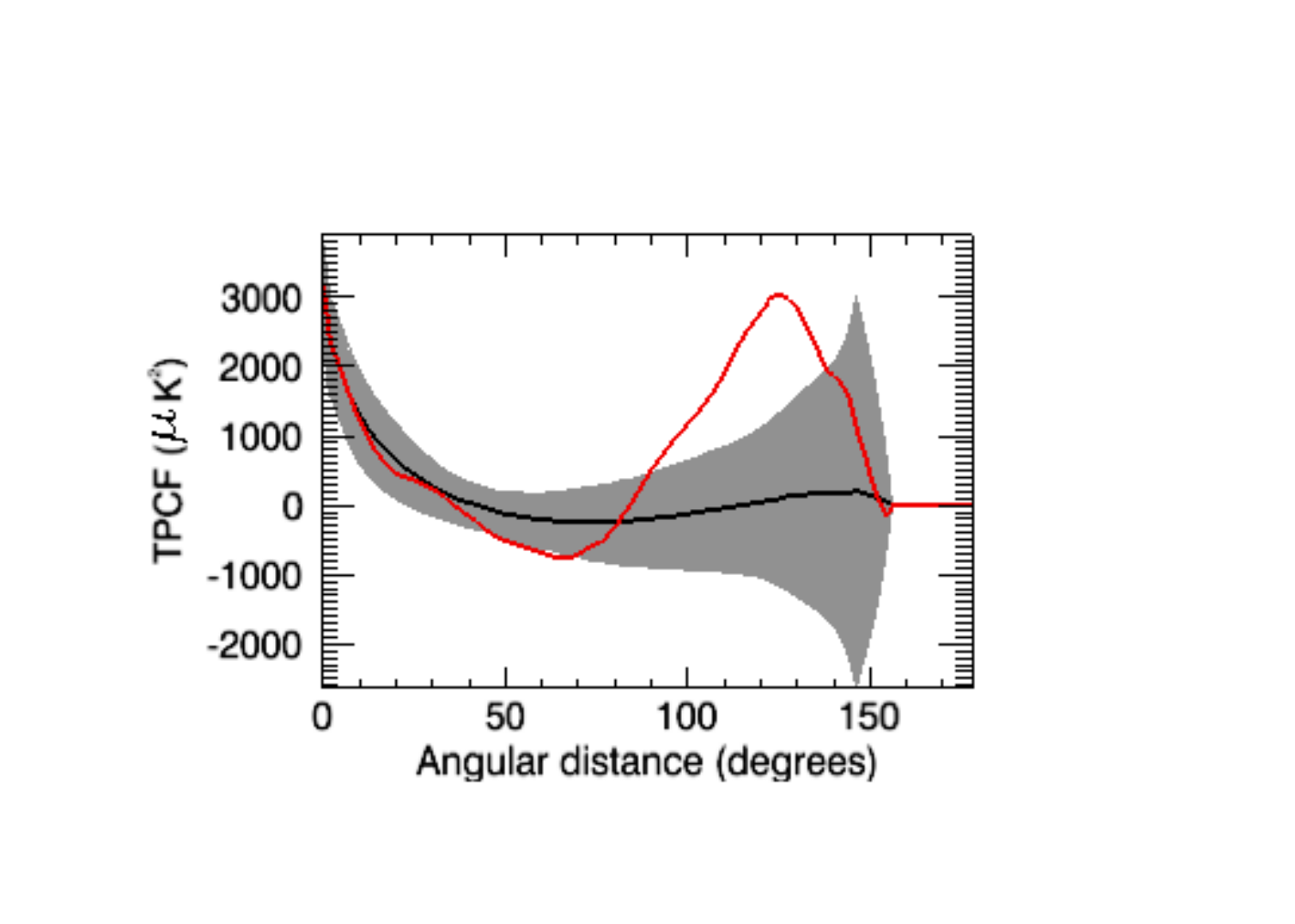}
\caption{TPCF curves computed for SMICA2 (red curve) using U73.}
\label{TPCF-u73}
\end{figure}

\begin{figure}
 \includegraphics[scale=0.5]{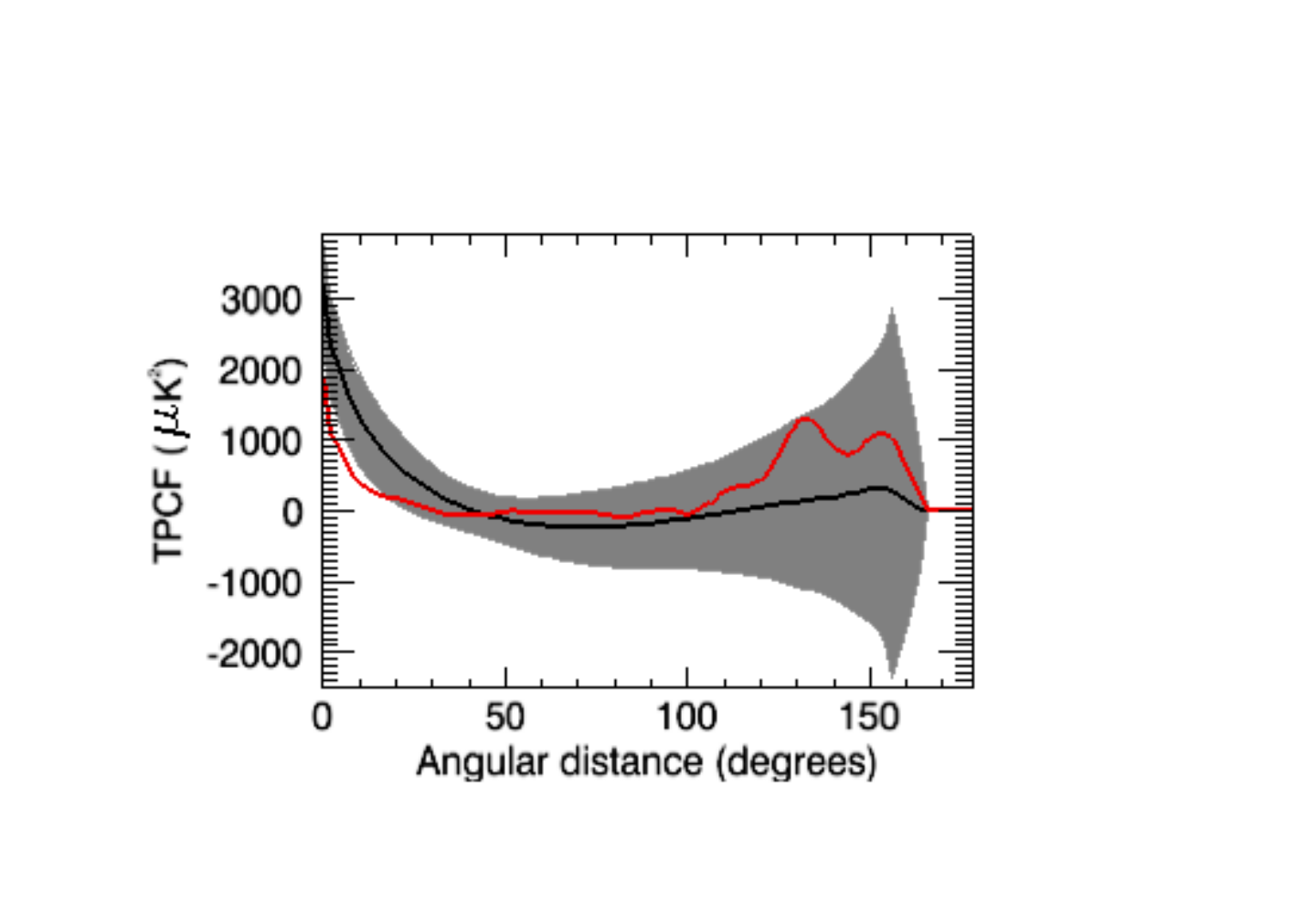}
 \includegraphics[scale=0.5]{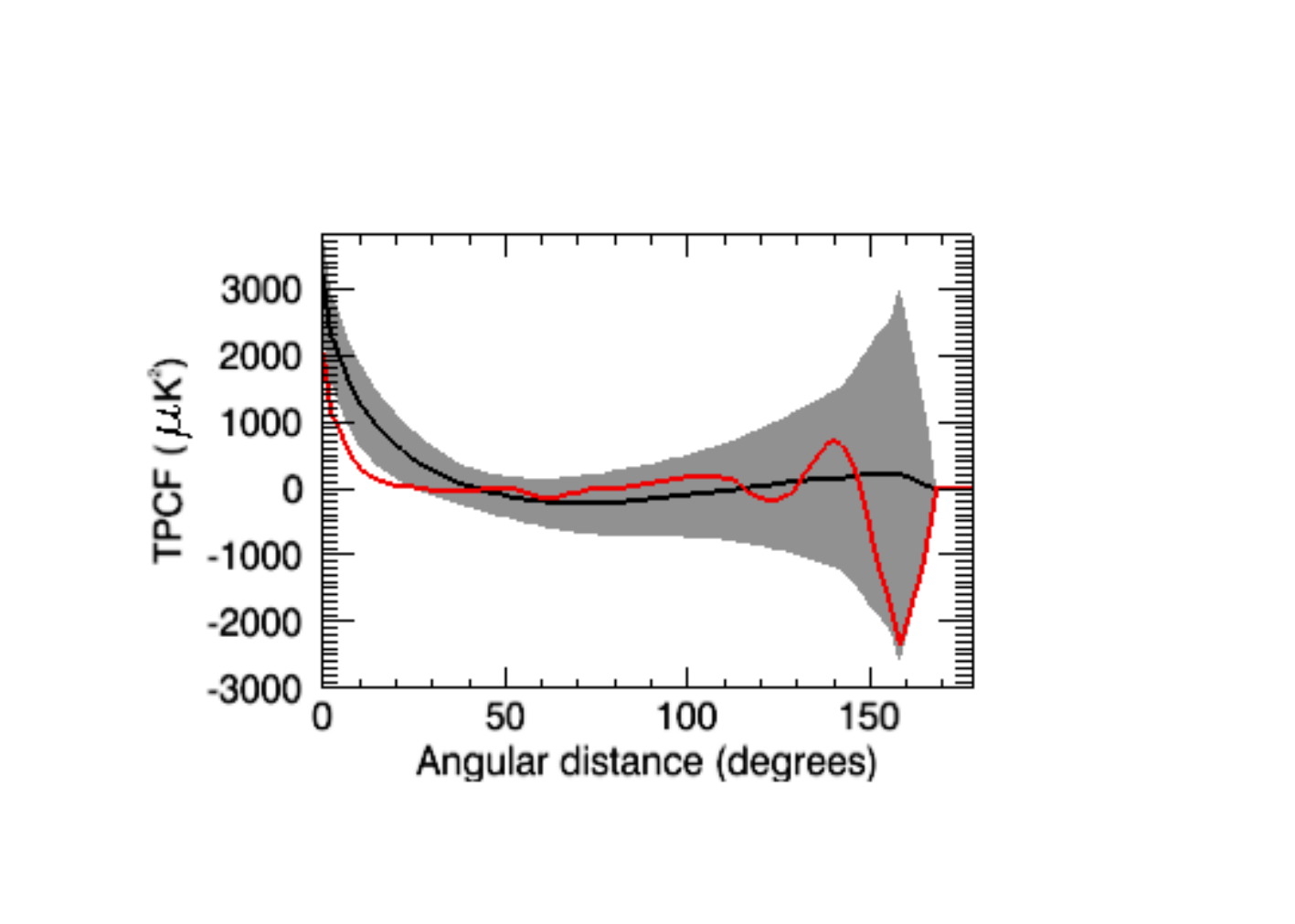}
 \includegraphics[scale=0.5]{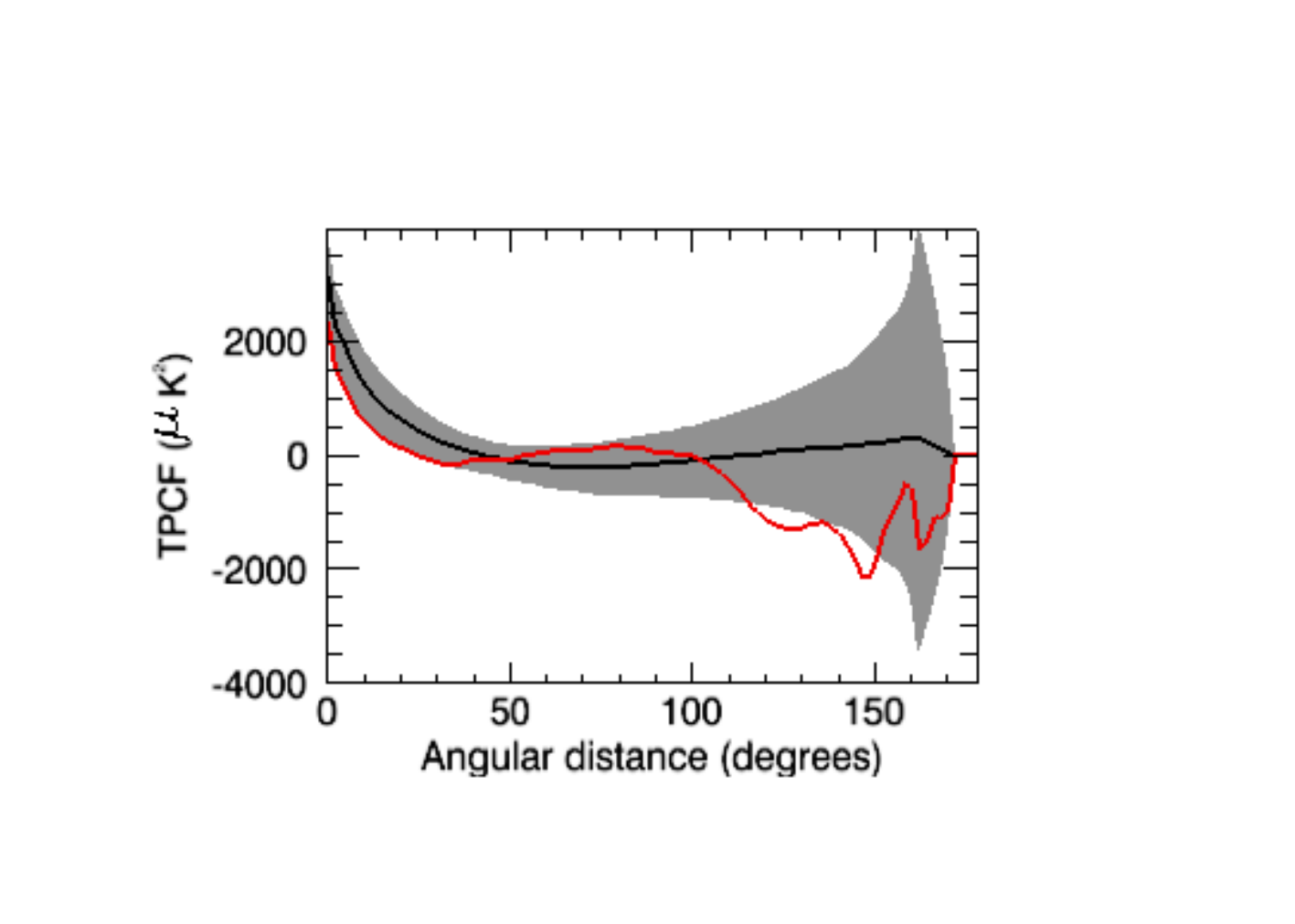}
 \includegraphics[scale=0.5]{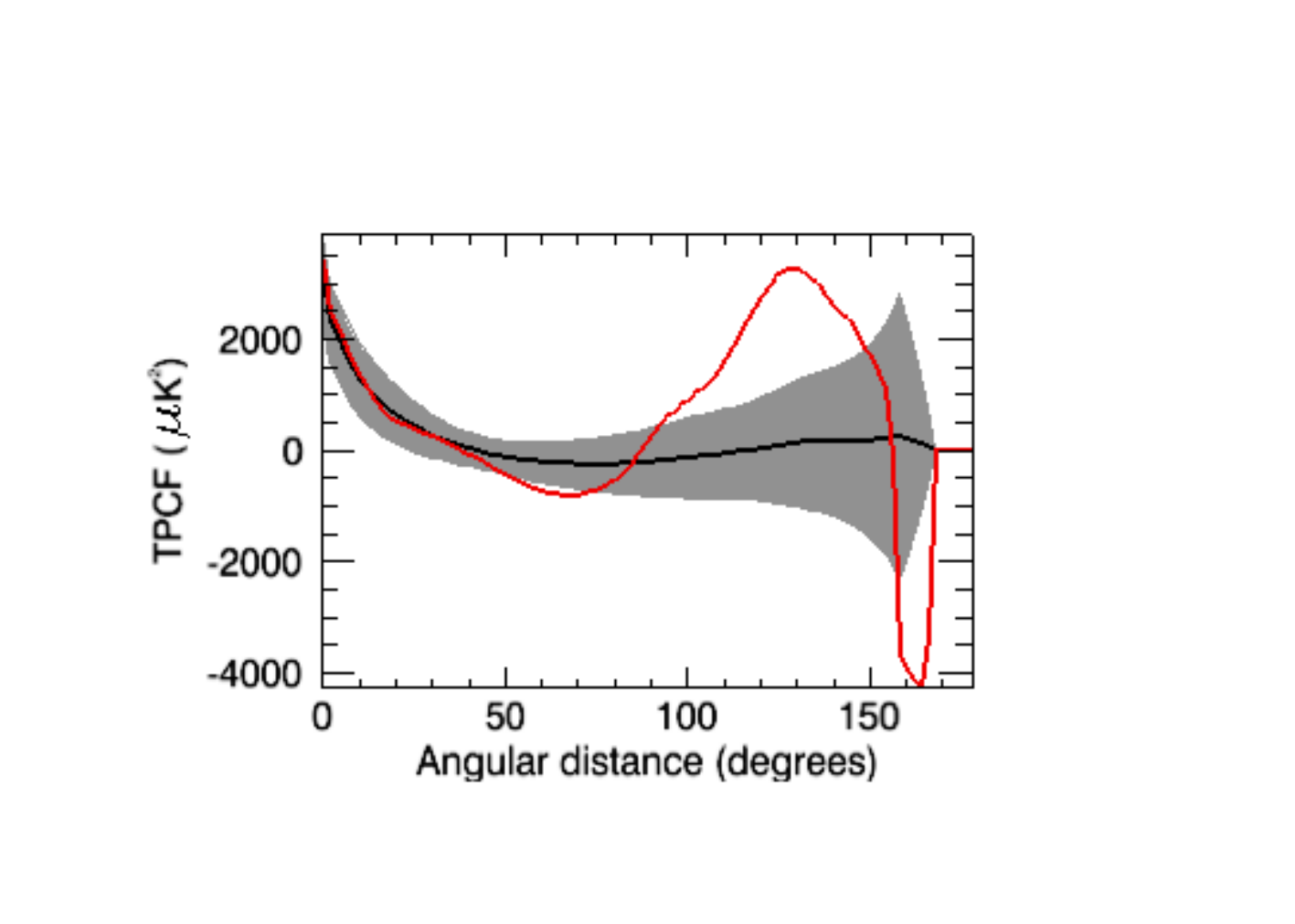}
\caption{TPCF curves computed for SMICA2 (red curve) using UT78}
\label{TPCF-UT78}
\end{figure}

\begin{figure}
 \includegraphics[scale=0.5]{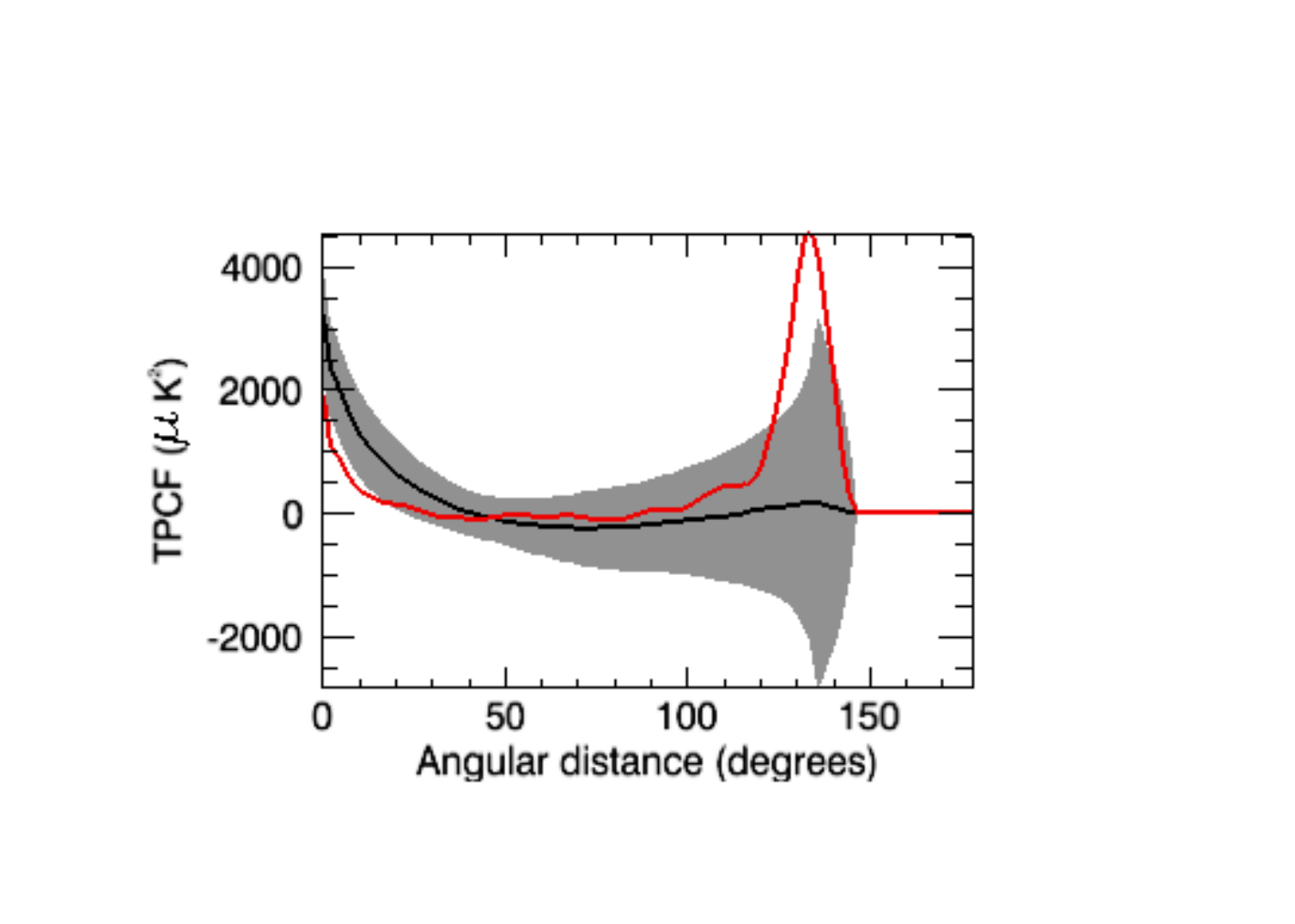}
 \includegraphics[scale=0.5]{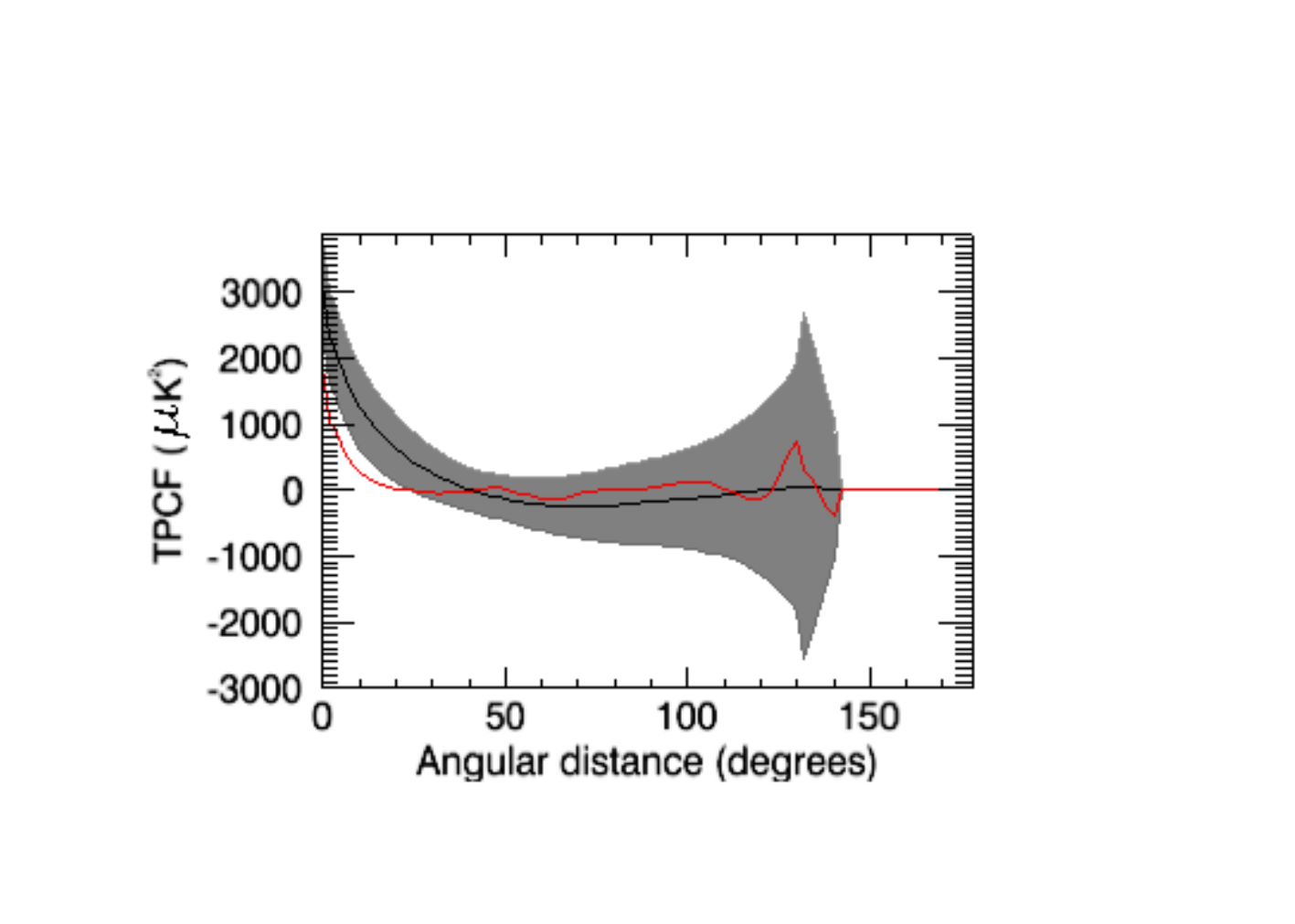}
 \includegraphics[scale=0.5]{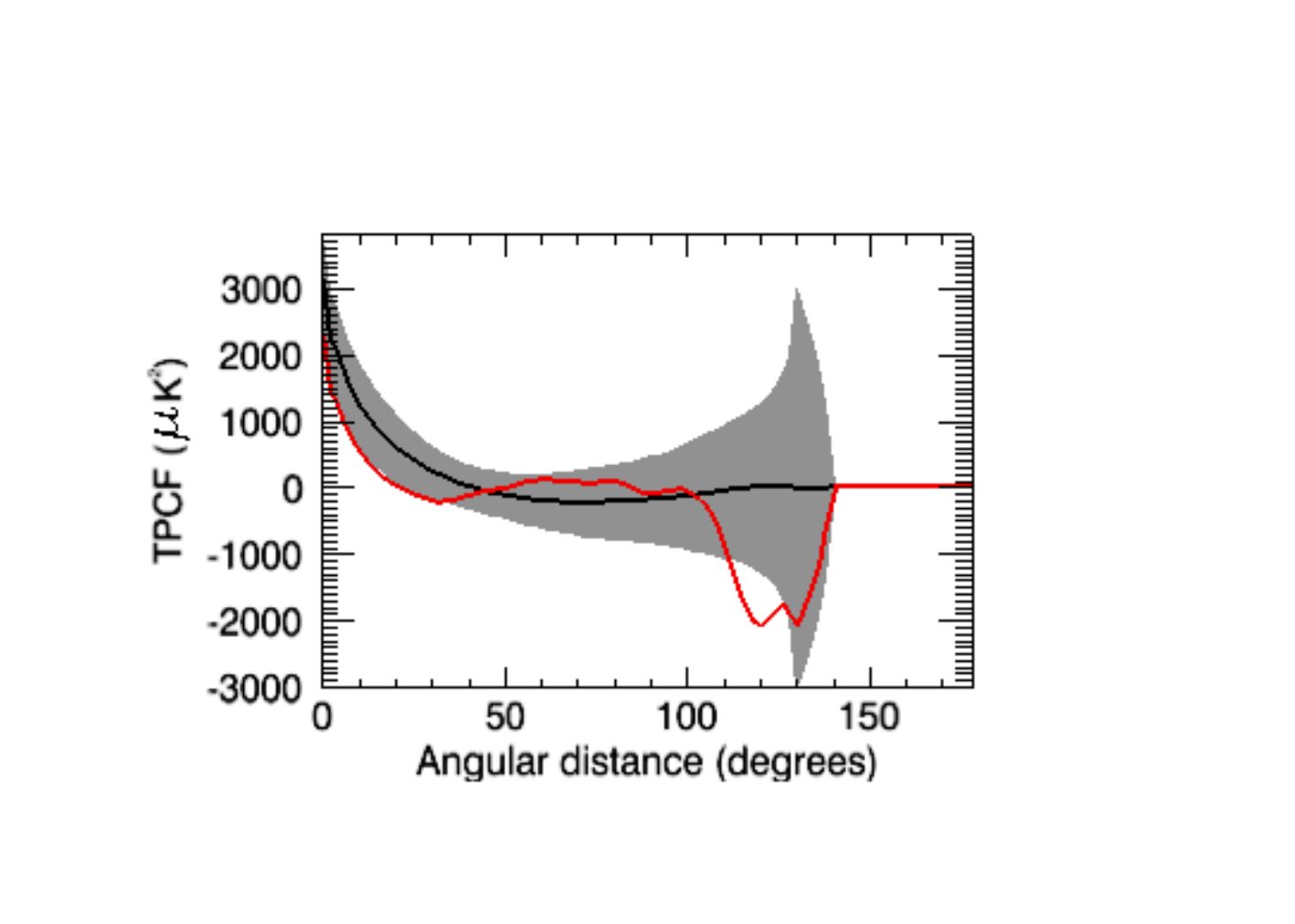}
 \includegraphics[scale=0.5]{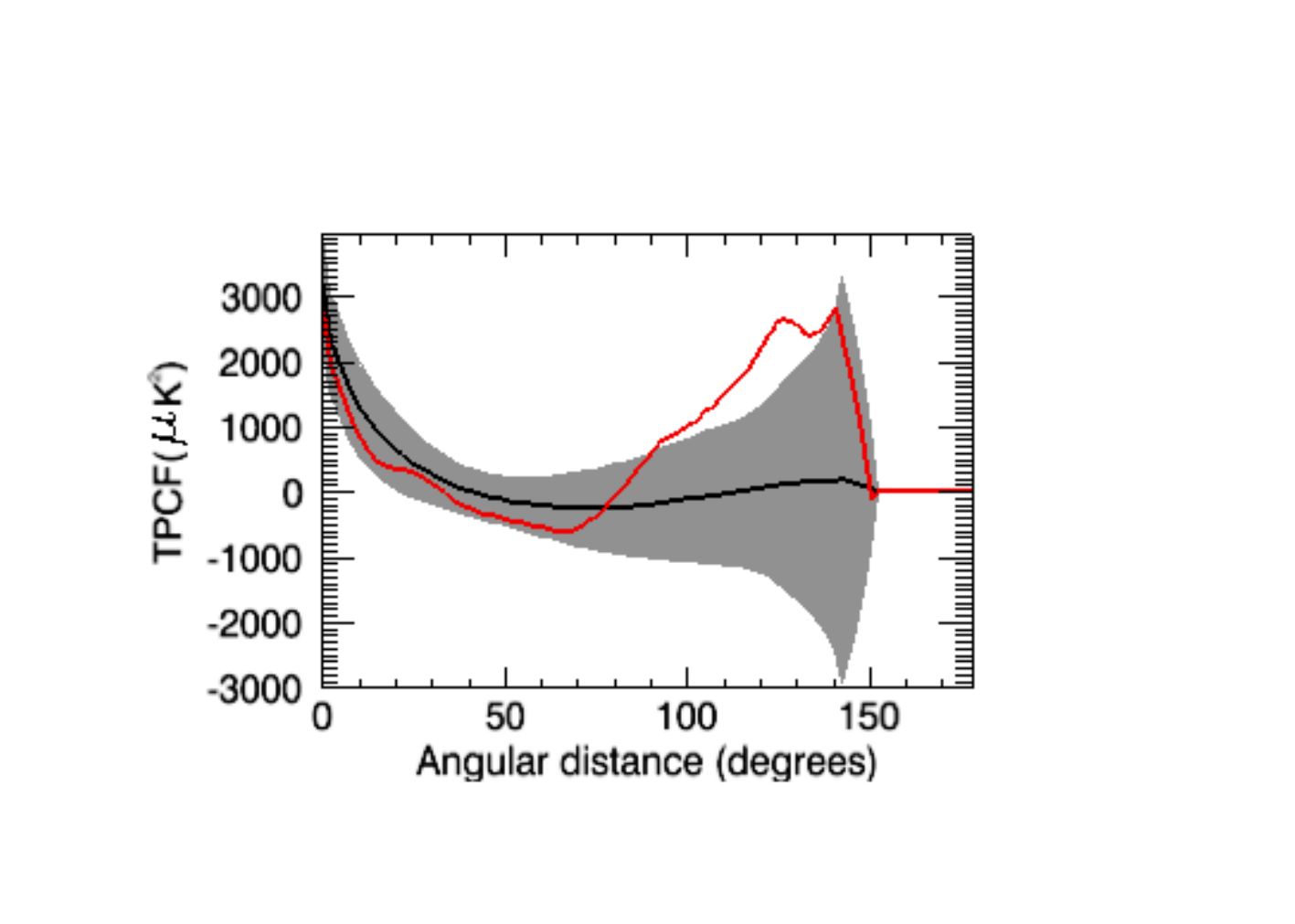}
\caption{TPCF curves computed for SMICA2 (red curve) using U66.}
\label{TPCF-U66}
\end{figure}

\begin{figure}
\includegraphics[scale=0.6]{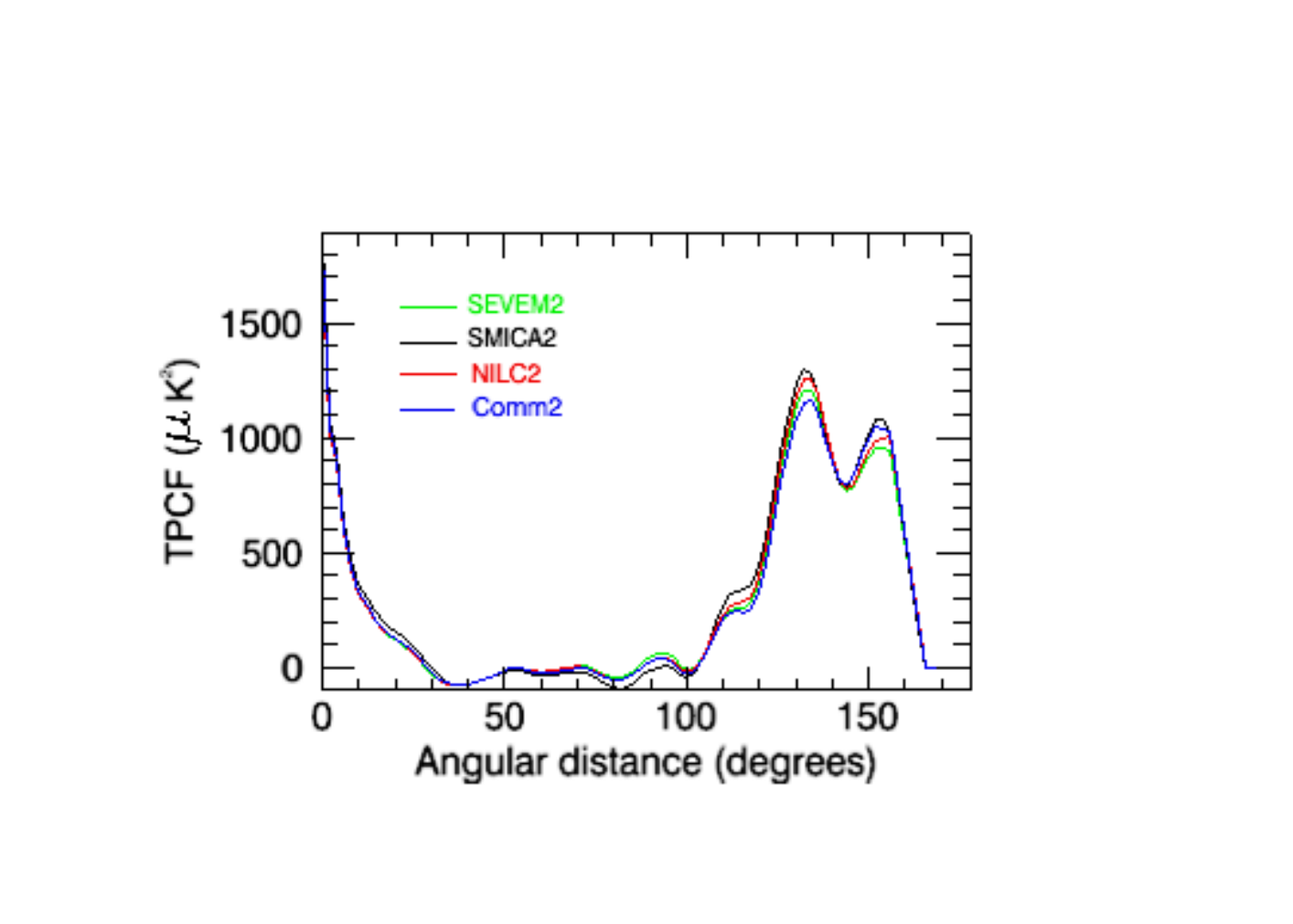}
\includegraphics[scale=0.6]{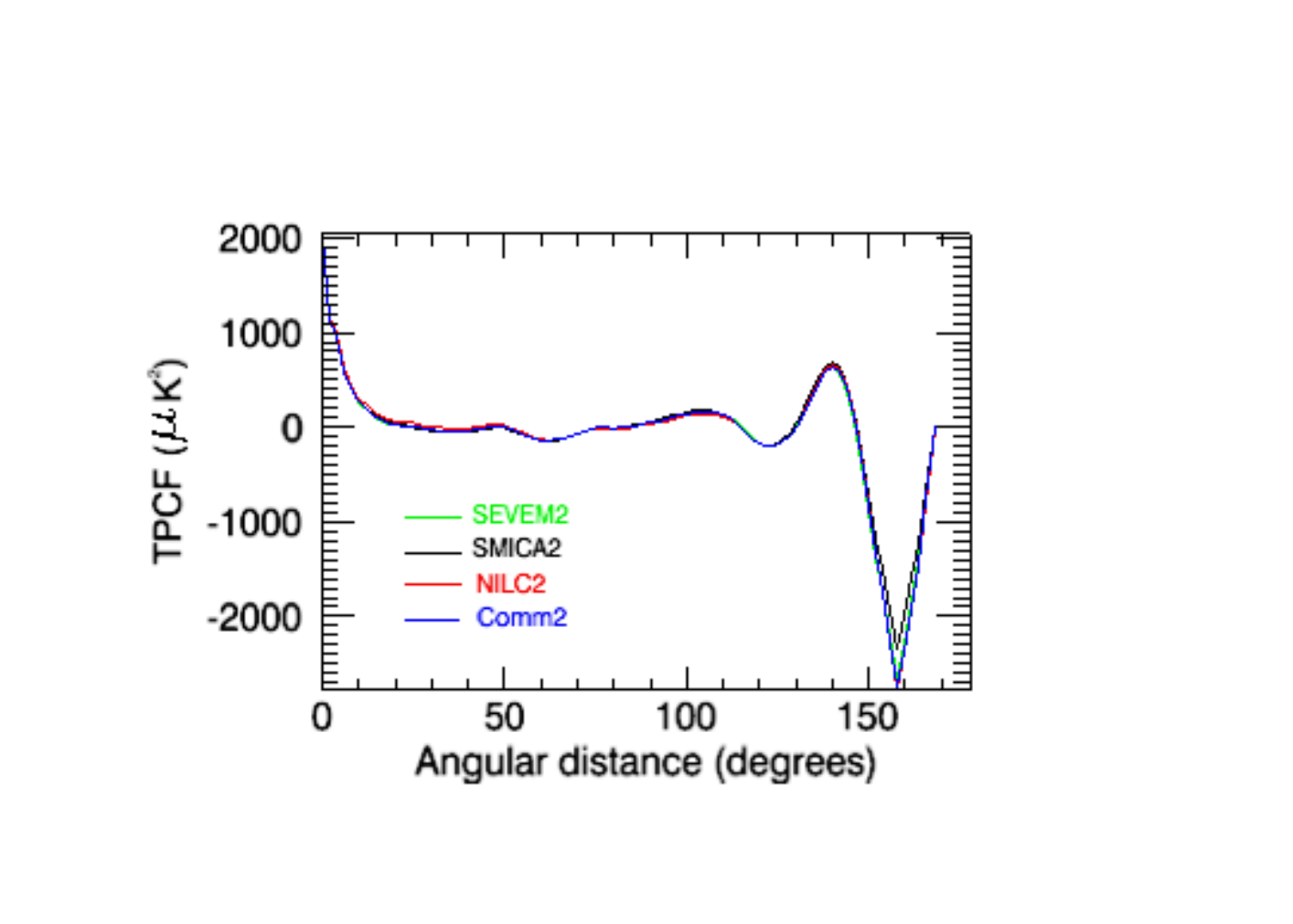}
\includegraphics[scale=0.6]{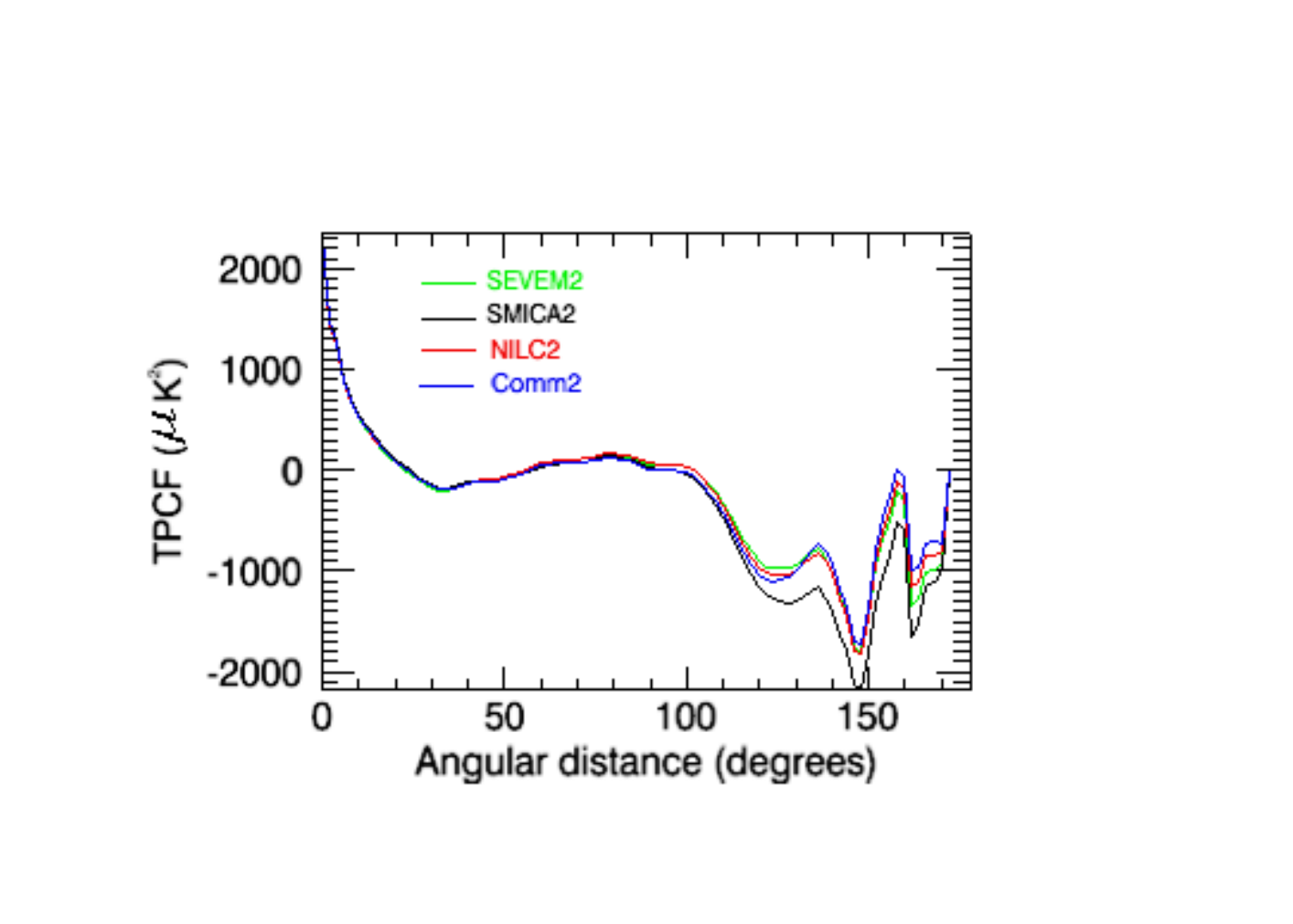}
\includegraphics[scale=0.6]{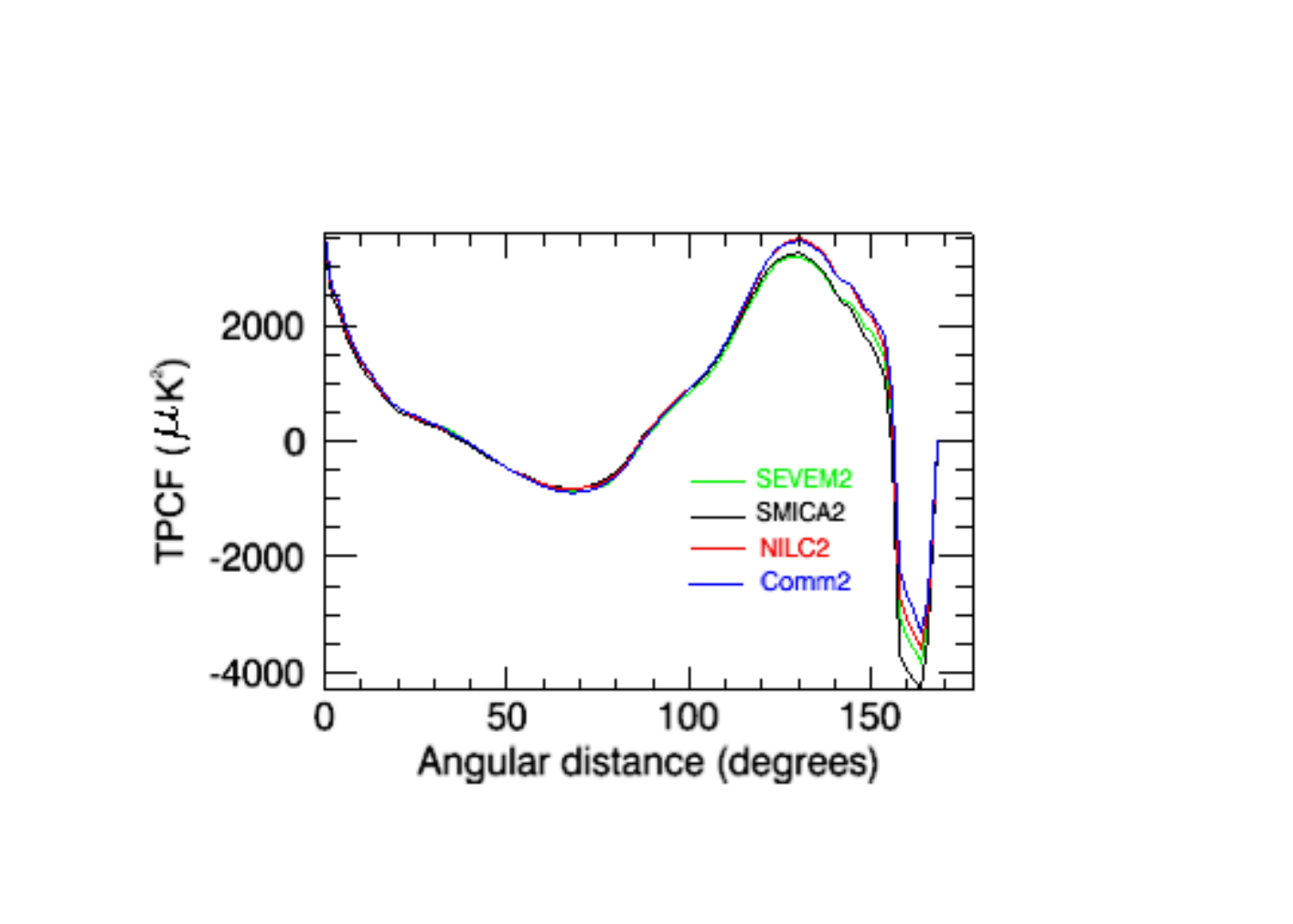}
\caption{Comparison between the Planck  foreground-cleaned maps using UT78 mask (SMICA2, NILC2, SEVEM2 and commander2). From  top to bottom, the curves refer to the NWQ, NEQ, SWQ, and SEQ.}
\label{TPCF-nilc}
\end{figure}

The TPCF computed for each quadrant of SMICA2 map is shown from Figures \ref{TPCF-rulerminimal}-\ref{TPCF-U66}, using masks mask-rulerminimal, U73, UT78 and U66, respectively. Using the quantity $\sigma$, defined in Equation \ref{sigma},  we quantify the results and compare them with the MC simulated maps for each mask described above. Moreover, we can see that the TPCF computed for the maps SMICA2, NILC2, SEVEM2 and Commander2  using mask UT78 agree to each other as it can be seen  in Figure \ref{TPCF-nilc}.

The analysis shows consistent results for three masks: mask-rulerminimal, U73 and U66 masks. For all these cases, we found a lack of large-angle temperature correlation in  the NEQ.  The probability that the exactly same quadrant in the simulations present the lack of correlation observed in SMICA2 is 0.2\% for both mask-rulerminimal and U73, and 1.4\% for U66. If we allow that at least one quadrant, among all four of them,  in each simulated map shares the absence of correlation presented in the data, we find values up to 5.4\% for U66. The excess of power in the SEQ also follows a pattern for these three masks: it tends to vanish as the number of excluded pixels gets larger, i.e., when the masks are more severe.  The probability that the excess of power occurs in the simulated maps varies, depending on which mask is being taken into account, from 8.2\% to 28.9\%, if we consider only the SEQ of the simulations. On the other hand, if we establish that at least one quadrant among all in every simulation can share the excess of power found in the data, the probability increases from 27.1\% to 64.1\%. For details see Tables \ref{tbl-prob-quadrants-exac} and \ref{tbl-prob-quadrants}.  

In short, we can say that even though there is an excess of power in the SEQ, it is in agreement with the $\Lambda$CDM model when we apply mask-rulerminimal, U73 and U66 masks to SMICA2.  However, we quantified the probability that both features, namely the lack of correlation in the NEQ and the excess of power in the SEQ, happen at the same time in CMB data as the asymmetry between both quadrants. We find that  $<0.1\%$, 1.4\% and 4.6\% of the simulated maps have the same asymmetry  as the SMICA2 CMB map for mask-rulerminimal, U73 and U66, respectively (see Table \ref{tbl-prob-mask}). Once more, if we allow that the asymmetry can happen between any pair of quadrants of a simulated map, the above  probabilities increase to 1.4\%, 6.3\% and 20.4\%, being considered no longer anomalous for conservative masks. For a comparison between the values of $\sigma_{SEQ}/\sigma_{NEQ}$ found in the data with the mean values, 68\% C.L., 95\% C.L and 99.7\% C.L found in the simulations see Table  \ref{sigma_simul}.

A different result was found for the new UT78 2015 Planck mask. In opposition to all other masks, when UT78 is applied to the data we found that the NEQ is no longer anomalous, meaning that now there is a large-angle temperature correlation in this quadrant, what makes it in agreement with the expect behavior defined by the standard cosmological model. On the other hand, only 2.3\% of the simulations have the excess of power in the SEQ found in SMICA2. In other words, the excess of power in the SEQ is the biggest when we use UT78. If any quadrant in the simulation can account for this excess of power the probability increases to 10.9\% (see again Tables\ref{tbl-prob-quadrants-exac} and \ref{tbl-prob-quadrants}). The chance that the asymmetry between NEQ and SEQ occurs in the simulations for the exact same pair of quadrants is of 2.7\% and 17.1\% between any pair of quadrants in each simulated map. 

So, for every mask used we still find an anomalous asymmetry between the NEQ and the SEQ, taking into account the same pair of quadrants in the simulations (considering that in this case they have the same number of pixels of the CMB map). The problem here is that the reason we find the asymmetry is different for UT78 when we compare to mask-rulerminimal, U73 and UT78. To investigate this issue, we combine UT78 with each one of the other masks and calculate the TPCF and its correspondent $\sigma$ value for each quadrant of the SMICA2 map. The results can be seen in Table \ref{tbl-prob-quadrants}. It is possible to notice that the new results are in agreement with the previous ones presented in this paper for mask-rulerminimal, U73 and U66 alone. The lack of correlation in the NEQ is still present in the data for UT78 combined with any of the other masks. The excess of power in the SEQ also follows the pattern described previously: it decreases as the mask becomes more severe. 

Furthermore, for completeness, we calculated the TPCF and the correspondent $\sigma$ value for SMICA2 map performing simple symmetric Galactic cuts of 10$^\circ$, 20$^\circ$, 25$^\circ$ and  30$^\circ$. We again found that the excess of power in the SEQ decreases as the Galactic cut gets more severe. Moreover, for a more severe symmetric cut of 30$^\circ$, the result agrees with the ones for mask-rulerminimal, U73 and U66 with a significant lack of power in the NEQ (see Table \ref{tbl-symmetric} for details). We also added to these symmetric Galactic cuts the point source masks provided by the Planck team in the first and second data release, obtaining no significant change in the results as presented in Table \ref{tbl-symmetric}.


Finally, to check if the results obtained for UT78 could be due to non-removed foregrounds, we calculated histograms and their statistics (skewness and kurtosis) for the regions in the sky not covered by UT78 but covered by mask-rulerminimal or U73 or U66. We compare the values obtained for both quantities using SMICA2 and the simulations. We found interesting results specially for the NEQ: the value for the kurtosis in these regions is always above 3 for the data and in average not bigger than 2.42 for the simulations as can be seen in Table \ref{tbl-kurt}. The value of the kurtosis for the regions uncovered by UT78, but covered by U73 in the NEQ (called UT78-U73), is 4.94, which is higher  when compared with the average value of 2.21 for the simulations. We did not find any simulated map among 1000 simulations with such a high value of kurtosis in the same region of the sky. This result means that CMB temperature distribution in the analyzed regions is highly concentrated around the mean if compared to the $\Lambda$CDM simulations. 

\begin{figure}
\includegraphics[scale=0.6]{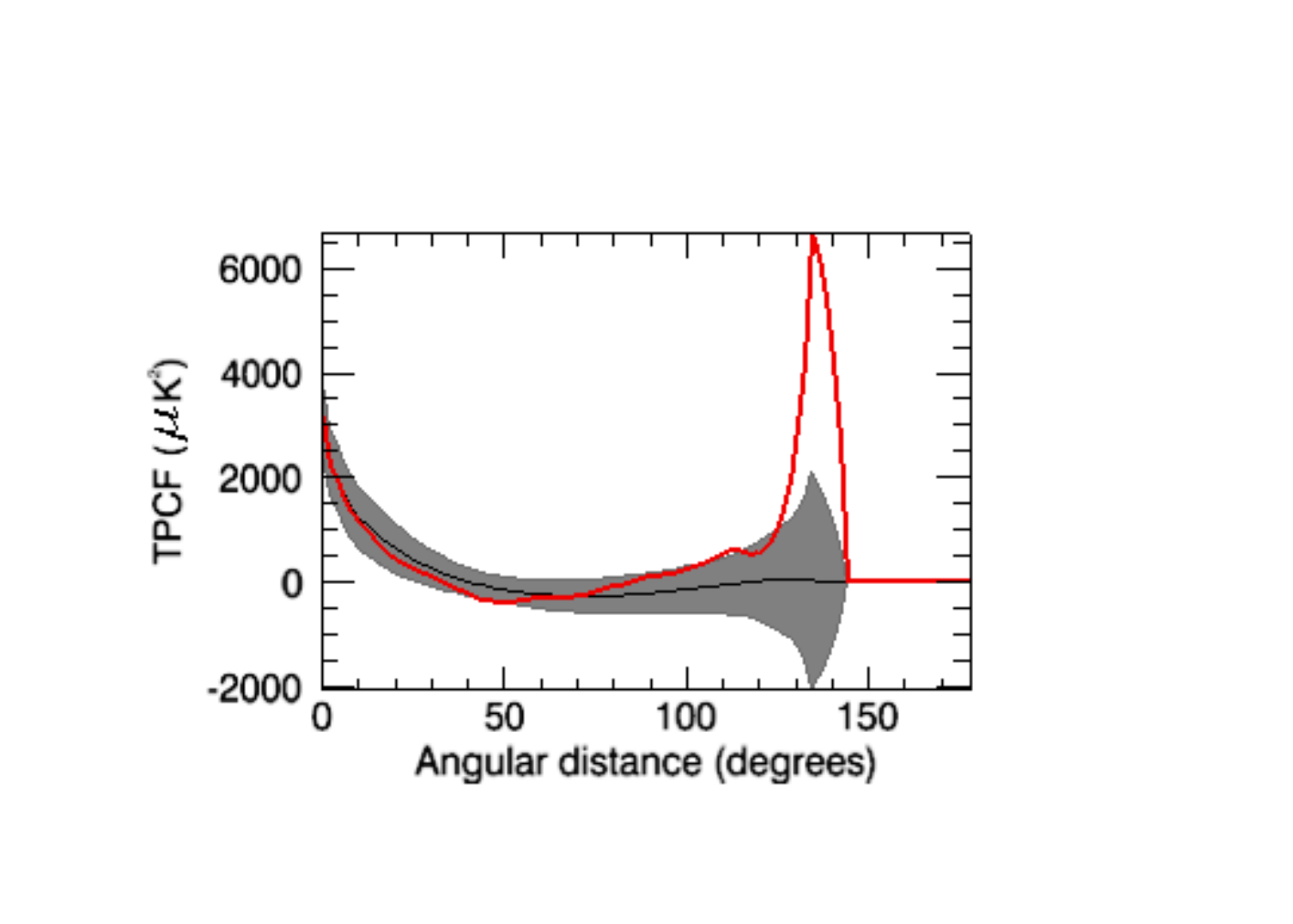}
\includegraphics[scale=0.6]{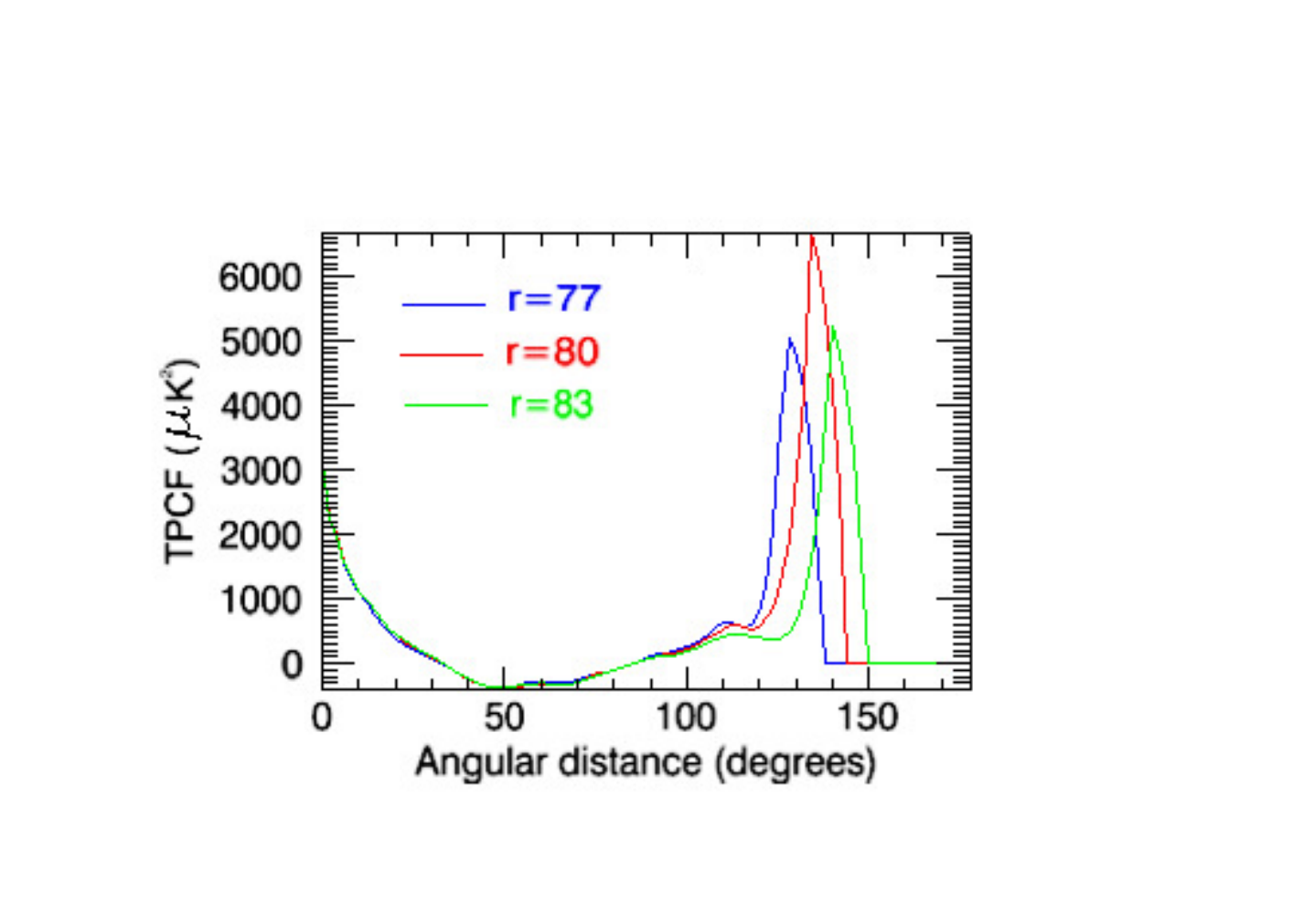}
\caption{Top: the TPCF curve computed for SMICA2 (red curve) using mask-rulerminimal in a circular region centered at $(\phi, \theta)=(270^{\circ}, 135^{\circ})$ and radius, $r=80^{\circ}$. The shadow part depicts the standard deviation intervals  (68\% C.L)  for 1000 simulated maps produced with the $\Lambda$CDM spectrum. The black curve is the mean TPCF considering the MC simulated maps. Bottom: the TPCF computed for SMICA2 using mask-rulerminimal in a circular region centered at $(\phi, \theta)=(270^{\circ}, 135^{\circ})$ and radius, $r=77^{\circ}, 80^{\circ}$ and $83^{\circ}$}
\label{TPCF-circ-SEQ}
\end{figure}

In addition to the previous analysis, we calculated the TPCF where we found the excess and lack of power in CMB sky using circular regions and compared with the results for the simulated sky maps. We found that the biggest excess of power falls in the region centered at $(\phi, \theta)=(270^{\circ}, 135^{\circ})$ (from now on region 1) (HEALPix convention)  and radius, $r=80^{\circ}$ (11.488 pixels available). The results can be seen in Figure \ref{TPCF-circ-SEQ}. In Figure \ref{TPCF-circ-SEQ}, we also compare the result for region 1 for $r=80^{\circ}$, $r=77^{\circ}$ and $r=83^{\circ}$. The excess of power tends to decrease for radius different than $r=80^{\circ}$ (check Table \ref{tbl-sigma-circ} for the correspondent $\sigma$ values for the TPCF in region 1 using $r=80^{\circ}$, $r=77^{\circ}$ and $r=83^{\circ}$).

\begin{figure}
\includegraphics[scale=0.6]{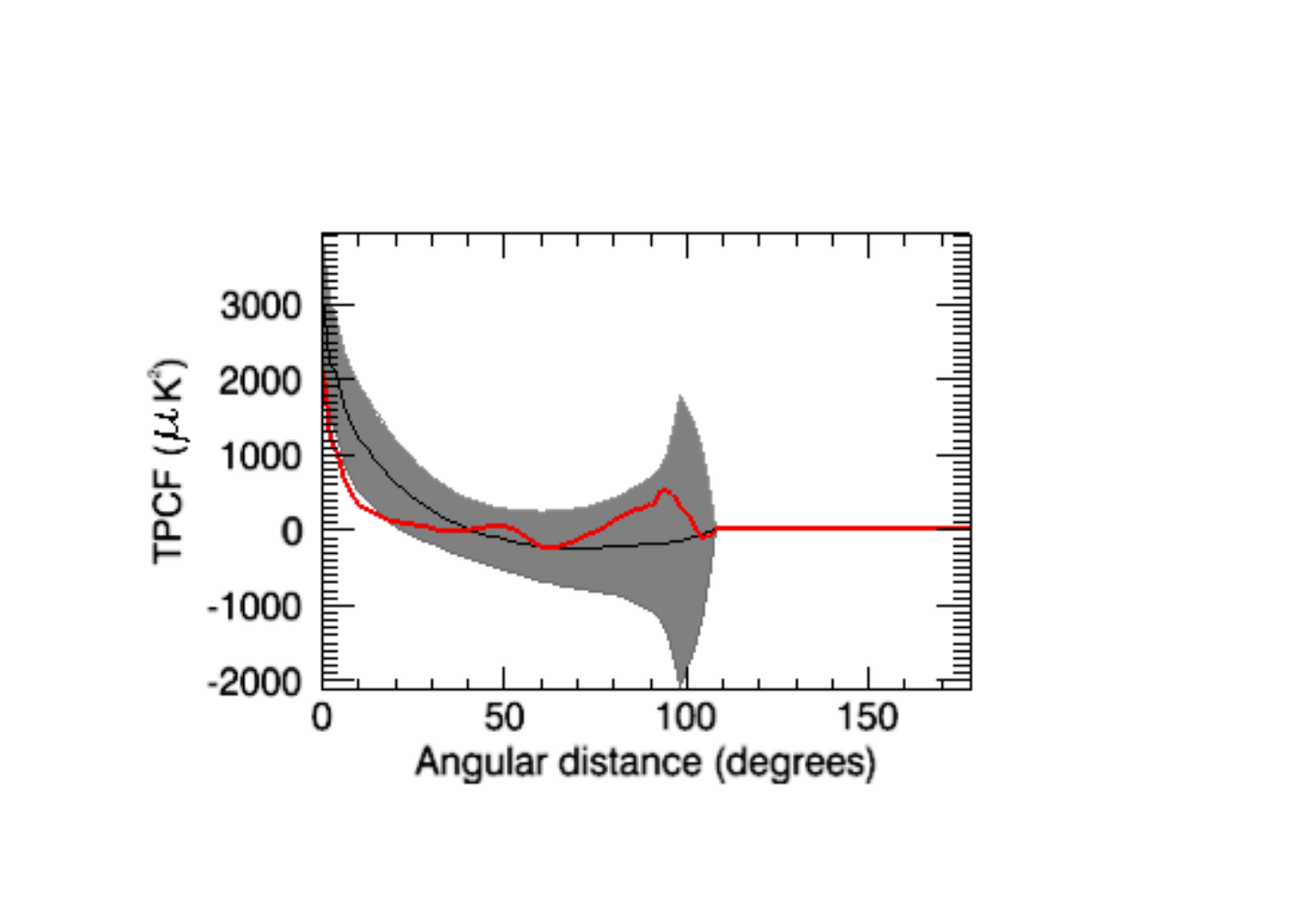}
\includegraphics[scale=0.6]{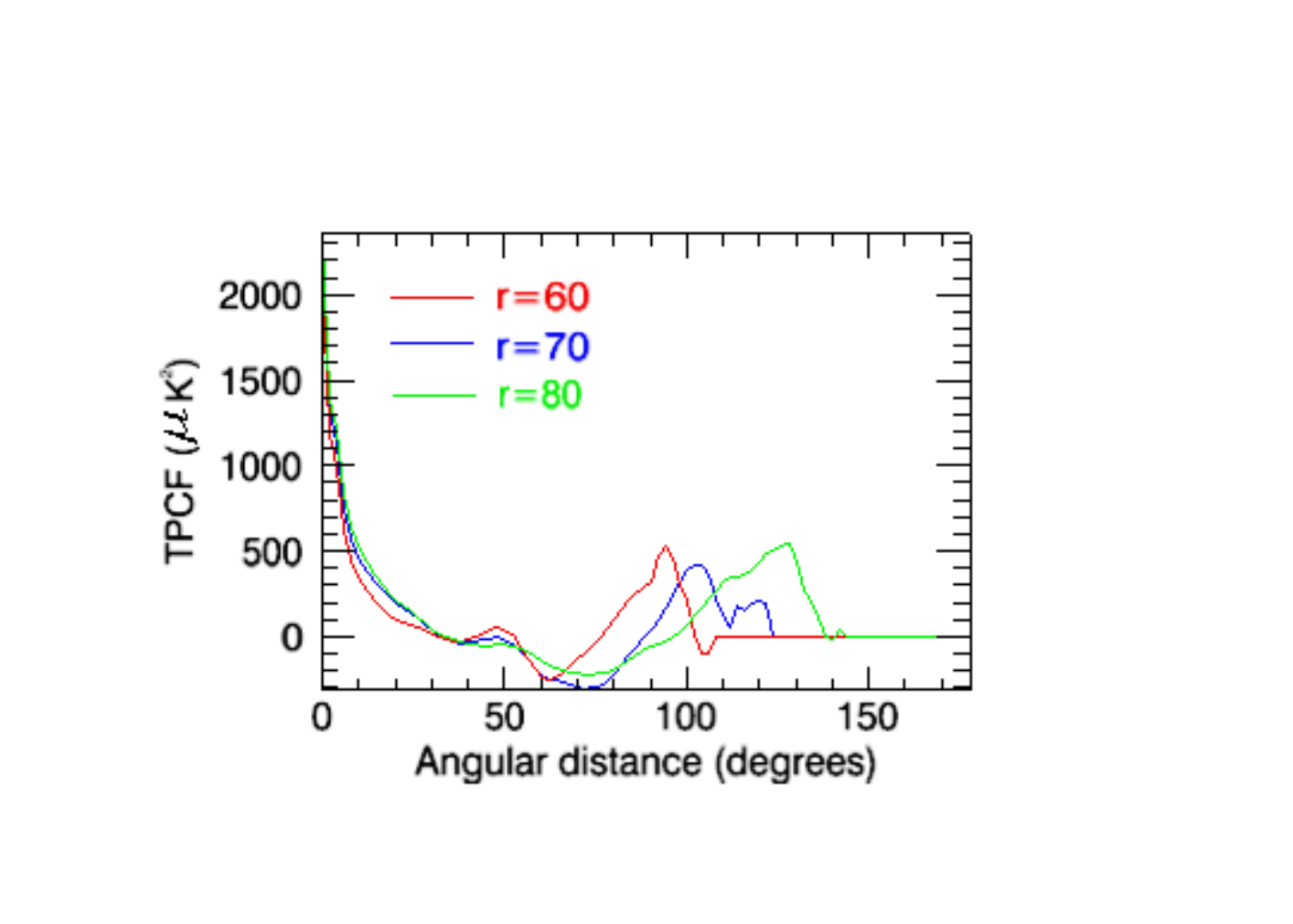}
\caption{Top: the TPCF curve computed for SMICA2 (red curve) using mask-rulerminimal in a circular region centered at $(\phi, \theta)=(270^{\circ}, 45^{\circ})$ and radius, $r=60^{\circ}$. The shadow part depicts the standard deviation intervals  (68\% C.L)  for 1000 simulated maps produced with the $\Lambda$CDM spectrum. The black curve is the mean TPCF considering the MC simulated maps. Bottom: the TPCF computed for SMICA2 using mask-rulerminimal in a circular region centered at $(\phi, \theta)=(270^{\circ}, 45^{\circ})$ and radius, $r=60^{\circ}, 70^{\circ}$ and $80^{\circ}$}
\label{TPCF-circ-NEQ1}
\end{figure}

\begin{figure}
\includegraphics[scale=0.6]{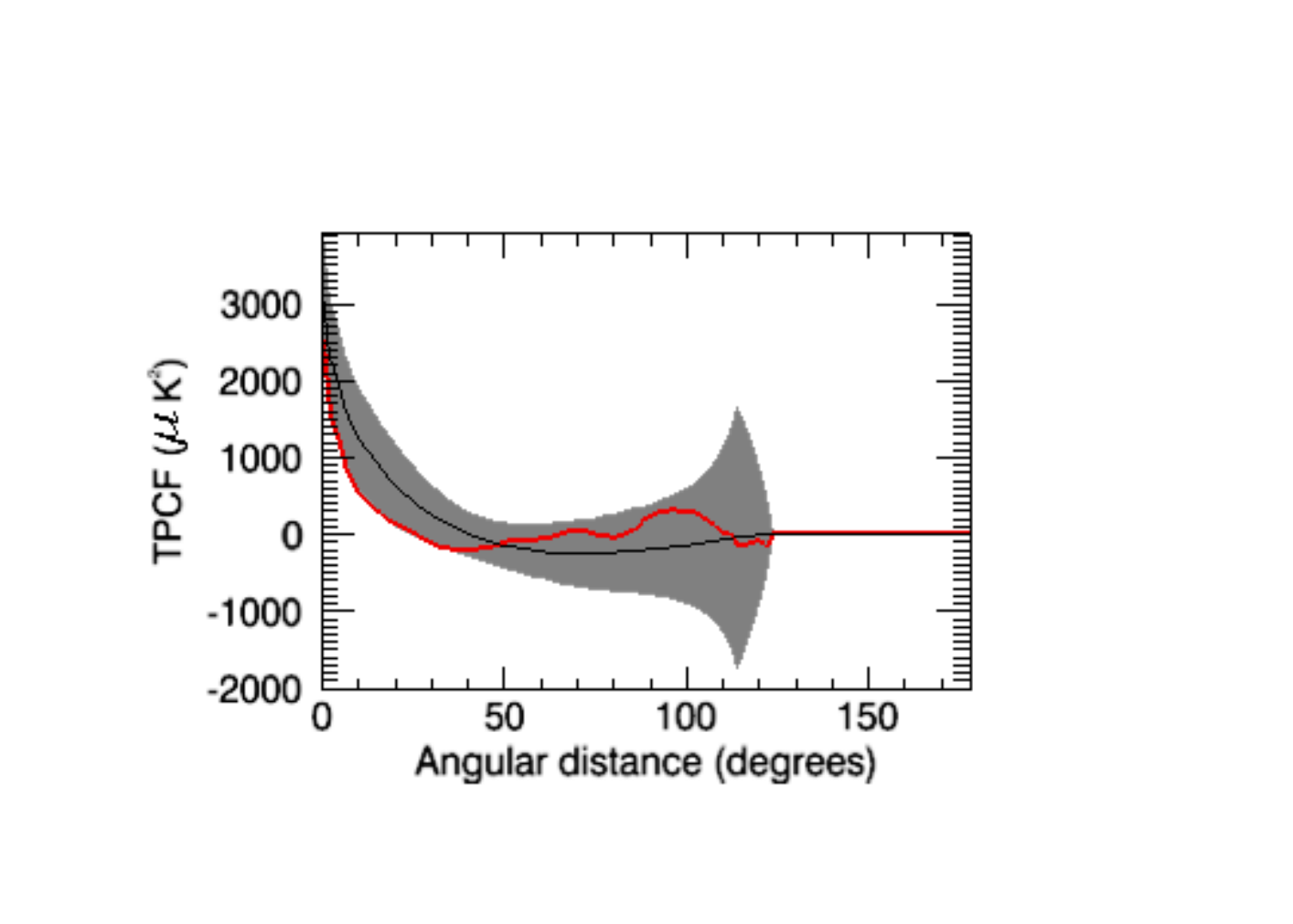}
\includegraphics[scale=0.6]{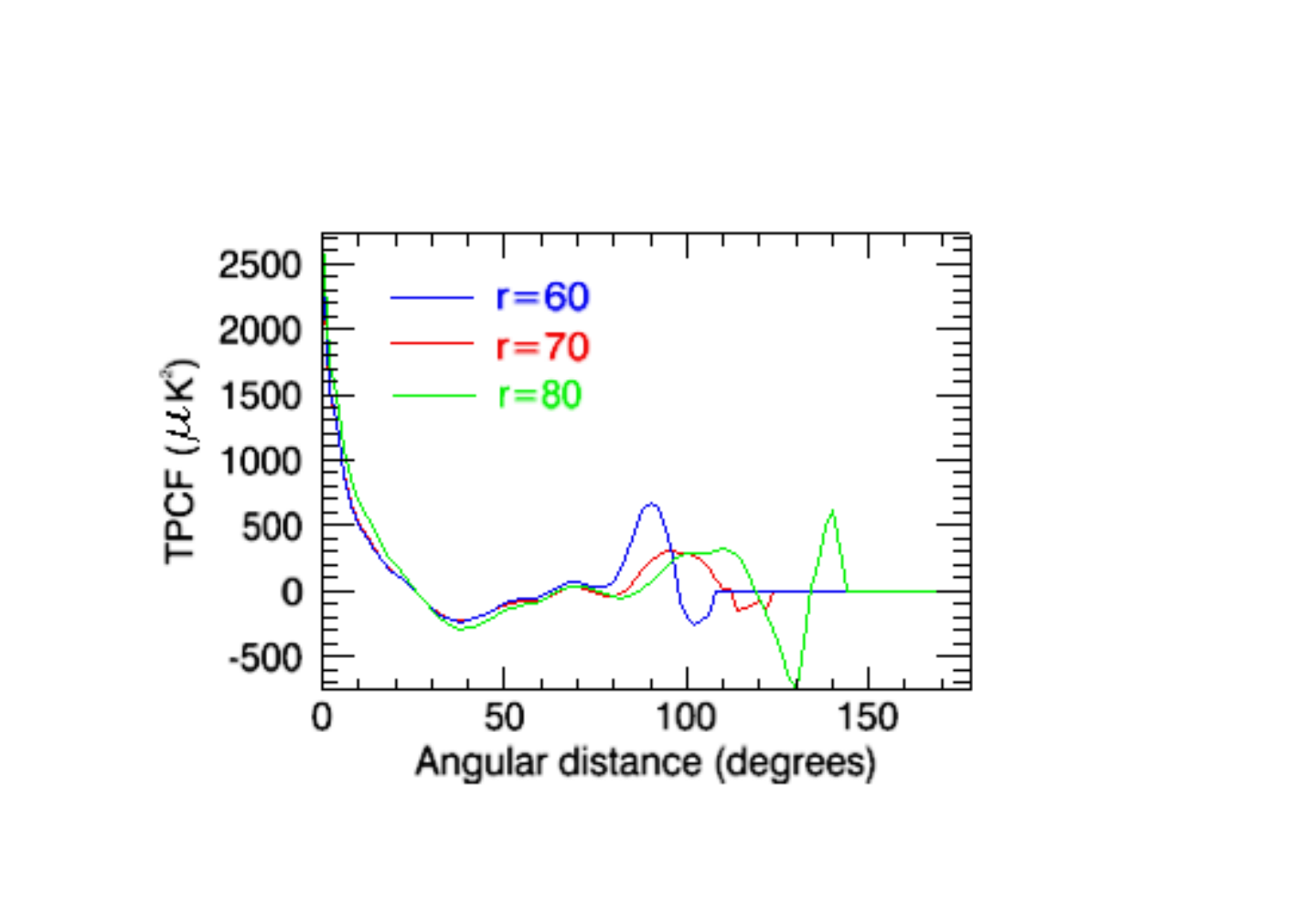}
\caption{Top: the TPCF curve computed for SMICA2 (red curve) using mask-rulerminimal in a circular region centered at $(\phi, \theta)=(225^{\circ}, 45^{\circ})$ and radius, $r=70^{\circ}$. The shadow part depicts the standard deviation intervals  (68\% C.L)  for 1000 simulated maps produced with the $\Lambda$CDM spectrum. The black curve is the mean TPCF considering the MC simulated maps. Bottom: the TPCF computed for SMICA2 using mask-rulerminimal in a circular region centered at $(\phi, \theta)=(225^{\circ}, 45^{\circ})$ and radius, $r=60^{\circ}, 70^{\circ}$ and $80^{\circ}$}
\label{TPCF-circ-NEQ2}
\end{figure}

In the same way, we found significant lack of power for regions centered at $(\phi, \theta)=(270^{\circ}, 45^{\circ})$ (region 2) and $(\phi, \theta)=(225^{\circ}, 45^{\circ})$ (region 3) for radius, r=60 (6988 pixels) and r=70 (9307 pixels), respectively (see Figures \ref{TPCF-circ-NEQ1} and \ref{TPCF-circ-NEQ2} for comparison of the TPCF in SMICA2 and simulations. If we considere region 2, we found that if we vary the radius from $60^{\circ}$ to $80^{\circ}$, the lack of power in this region remains nearly unchanged (see Table \ref{tbl-sigma-circ}).  

We found that both the excess of power in region 1($80^{\circ}$) and lack of power in regions 2 ($60^{\circ}$)  and 3 ($70^{\circ}$) have low probability to occur in the simulations, being 1\%, 2.1\% and 5.3\%, respectively. These later results are in agreement with the previous ones (when we divided the sky in quadrants) as expected. We can compare the results for the quadrants and for the new circular regions by looking at Table \ref{tbl-prob-quadrants-exac} (first line, for mask-rulerminimal) and  Table \ref{tbl-prob-circ-exac}.

\section{Conclusion}

We found a good agreement among the results when considering mask-rulerminimal, U73 and U66 in the analysis of CMB angular distribution. For all these masks, we confirm previous results in which CMB data present a significant lack of large-angle temperature correlation in the NEQ, which is anomalous if we assume the $\Lambda$CDM. In the same way, we found an excess of power in SEQ that decreases as the mask becomes more severe. Nevertheless, the asymmetry between the SEQ and the NEQ is confirmed in CMB data for mask-rulerminimal, U73 and U66 with at least 95\% C.L.  The excess of power in the SEQ computed when applying a symmetric cut around the Galactic plane is also in agreement with mask-rulerminimal, U73 and U66. The lack of power in the NEQ is also in agreement with the cited masks for a conservative symmetric cut of 30$^\circ$. 

A different result appears in the analysis for the new Planck mask released by Planck team in 2015. In this case, even though the asymmetry between the SEQ and NEQ is still confirmed in the data, the NEQ is in agreement with the $\Lambda$CMD model and the SEQ presents the biggest excess of power among all the other tested masks. Further calculations show that regions in the sky, specially in the NEQ, uncovered by UT78 but covered by other masks, have a high value of kurtosis in the data, in disagreement with the values found in the simulations. This result suggests that UT78 leaves residual foregrounds in the data unmasked, if so, being unsuitable for CMB cosmological analysis.

 Finally, as expected,  we conclude by choosing different regions in the the CMB sky that the power asymmetry is not dependent of our previous choice of quadrants.
 
The authors acknowledge the use of HEALPix packages, of the Legacy Archive for Microwave Background Data Analysis (LAMBDA) and  of the Planck Legacy Archive (PLA).  The development of Planck has been supported by: ESA; CNES, and CNRS/INSU-IN2P3-INP (France); ASI, CNR, and INAF (Italy); NASA and DoE (USA); STFC and UKSA (UK); CSIC, MICINN, and JA (Spain); Tekes, AoF, and CSC (Finland); DLR and MPG (Germany); CSA (Canada); DTU Space (Denmark); SER/SSO (Switzerland); RCN (Norway); SFI (Ireland); FCT/MCTES (Portugal); and the development of Planck has been supported by: ESA; CNES, and CNRS/INSU-IN2P3-INP (France); ASI, CNR, and INAF (Italy); NASA and DoE (USA); STFC and UKSA (UK); CSIC, MICINN, and JA (Spain); Tekes, AoF, and CSC (Finland); DLR and MPG (Germany); CSA (Canada); DTU Space (Denmark); SER/SSO (Switzerland); RCN (Norway); SFI (Ireland); FCT/MCTES (Portugal); and PRACE (EU). A description of the Planck Collaboration and a list of its members, including the technical or scientific activities in which they have been involved, can be found at http://www.sciops.esa.int/index.php?project=planck\&page\\
=Planck\_Collaboration. 

W. Zhao and L. Santos are supported by project 973 under Grant No. 2012CB821804, 
by the NSFC No. 11173021, No. 11322324 and No.11421303, and by a project 
of KIP of CAS. T. Villela acknowledges the support of CNPq through grant 308876/2014-8.

\clearpage
\onecolumn

\begin{table}
\begin{center}
\caption{Probabilities of finding in exactly the same quadrant of the simulated MC maps bigger or smaller  $\sigma$ values than in Planck SMICA2 map. We show the effect in each quadrant of different masks provided by the Planck team.}
\label{tbl-prob-quadrants-exac} 
\begin{tabular}{ccrcrcrcr}
\hline\hline
 Map &  $\sigma_{SEQ}$ & P1 \tablefootmark{a} & $\sigma_{SWQ}$ & P2\tablefootmark{b} & $\sigma_{NEQ}$ & P3\tablefootmark{c}& $\sigma_{NWQ}$ & P4\tablefootmark{d} \\
  \hline
 SMICA2 +  mask-rulerminimal  &  1891.99 & 8.2\%    &506.74    &10.8\%    & 308.32  &  0.2\% & 420.39 & 2.4\% \\ 
 \hline
 SMICA2 +  U73  &  1356.25 & 20.3 \%    &567.98    & 16.2\%    & 302.18 & 0.2\% & 810.99 &54.3 \% \\ 
 \hline
 SMICA2 +  UT78  &  2373.69 & 2.3 \%    & 985.71    &46.4 \%    & 677.26 & 29.1 \% & 558.94 & 12.9\% \\ 
 \hline
   SMICA2 + U66    &1257.18 & 28.9\%   &829.16 &45.2\%    & 374.49 &1.4\%   &1233.65 &24.3\% \\ 
 \hline
\end{tabular}
\end{center}
\tablefoottext{a}{Probability of finding $\sigma_{MC}>\sigma_{SEQ}$ }
\tablefoottext{b}{Probability of finding $\sigma_{MC}<\sigma_{SWQ}$ }
\tablefoottext{c}{Probability of finding $\sigma_{MC}<\sigma_{NEQ}$ }
\tablefoottext{d}{Probability of finding $\sigma_{MC}<\sigma_{NWQ}$. An exception was made for SMICA2 + U66 where the probability of finding $\sigma_{MC} >\sigma_{NWQ}$ was calculated instead.}
\end{table}

\begin{table}
\begin{center}
\caption{Probabilities of finding in at least one quadrant of the simulated MC maps bigger or smaller  $\sigma$ values than in Planck SMICA2 map. We show the effect in each quadrant of different masks provided by the Planck team.}
\label{tbl-prob-quadrants} 
\begin{tabular}{ccrcrcrcr}
\hline\hline
 Map &  $\sigma_{SEQ}$ & P1 \tablefootmark{a} & $\sigma_{SWQ}$ & P2\tablefootmark{b} & $\sigma_{NEQ}$ & P3\tablefootmark{c}& $\sigma_{NWQ}$ & P4\tablefootmark{d} \\
  \hline
 SMICA2 +  mask-rulerminimal  &  1891.99 & 27.1\%    &506.74    &25.3\%    & 308.32  &  0.6\% & 420.39 & 8.8\% \\ 
 \hline
 SMICA2 +  U73  &  1356.25 & 43.8 \%    &567.98    & 55.1\%    & 302.18 & 1 \% & 810.99 &88.6 \% \\ 
 \hline
 SMICA2 +  UT78  &  2373.69 & 10.9 \%    & 985.71    &94.6 \%    & 677.26 & 64.1 \% & 558.94 & 41.1\% \\ 
 \hline
   SMICA2 + U66    &1257.18 & 64.1\%   &829.16 &85.9\%    & 374.49 &5.4\%   &1233.65 &65.6\% \\ 
 \hline
\end{tabular}
\end{center}
\tablefoottext{a}{Probability of finding $\sigma_{MC}>\sigma_{SEQ}$ }
\tablefoottext{b}{Probability of finding $\sigma_{MC}<\sigma_{SWQ}$ }
\tablefoottext{c}{Probability of finding $\sigma_{MC}<\sigma_{NEQ}$ }
\tablefoottext{d}{Probability of finding $\sigma_{MC}<\sigma_{NWQ}$. An exception was made for SMICA2 + U66 where the probability of finding $\sigma_{MC} >\sigma_{NWQ}$ was calculated instead.}
\end{table}


 \begin{table}
\begin{center}
\caption{Calculated probabilities of finding the asymmetry between the SEQ and the NEQ equal to or higher than those found in Planck data in the MC simulations using the Planck masks described previously considering the $\Lambda$CDM model.}
\label{tbl-prob-mask} 
\begin{tabular}{ccrrr}
\hline\hline
 Map & $\sigma_{SEQ}/\sigma_{NEQ}$ & P1 \tablefootmark{a} &P2 \tablefootmark{b} & P3 \tablefootmark{c}\\
 \hline
 SMICA2 + mask-rulerminimal              &6.1   & $<$0.1\%    &0.3\%    &1.4\%  \\ 
 SMICA2 + U73                                     &4.5   &1.4\%          &2.8\%    &6.3\%  \\ 
 SMICA2 + UT78                                   &3.5   &2.7\%         &4.2\%   &17.1\% \\ 
 SMICA2 + U66                                     &3.4   &4.6\%          &9.7\%    &20.4\% \\ 
\hline\hline 

\end{tabular}
\end{center}
\tablefoottext{a}{Probability of finding the asymmetries in the simulations for exactly same configuration as in the SMICA map.}
\tablefoottext{b}{Probability of finding the asymmetries between the SEQ quadrant  and  any other quadrant in the simulations.}
\tablefoottext{c}{Probability of finding the asymmetries between any pair of quadrants in the simulations.}

\end{table}

\begin{table}
\begin{center}
\caption{Mean, 68.2\%, 95\%, and 99.7\% C.L.  values of  $\sigma_{SEQ}/\sigma_{NEQ}$  using the simulated CMB maps considering the $\Lambda$CDM model for the four studied masks. \label{tbl-sigma} }
\label{sigma_simul} 
\begin{tabular}{ccccc}
\hline\hline
   Masks  &mean &68\% C.L. &95\% C.L. & 99.7\% C.L.\\
 \hline                                                           

mask-rulerminimal &1.14    &1.13 &2.55 &4.45\\ 
\hline
U73 &1.33    &1.22 &2.85 &6.31\\ 
\hline
UT78 &1.21    &1.13 &2.82 &5.26\\ 
\hline
U66& 1.37   &1.38 &3.16 & 6.32 \\
\hline\hline 
   
\end{tabular}
\end{center}
\end{table}

\begin{table}
\begin{center}
\caption{$\sigma$ values for each quadrant in SMICA2 temperature map using combined Planck masks}
\label{tbl-prob-quadrants} 
\begin{tabular}{ccccc}
\hline\hline
 Combined masks &  $\sigma_{SEQ}$  & $\sigma_{SWQ}$& $\sigma_{NEQ}$ & $\sigma_{NWQ}$  \\
 \hline
   mask-rulerminimal  + UT78   &1585.87   &839.41     & 378.95   &536.62 \\ 
  \hline
  U73 + UT78                           &1404.45   &556.13      & 315.54  &795.92 \\ 
 \hline
   U66+ UT78                           &1336.02   &1019.29    & 379.08   &1221.22 \\ 
 \hline 
                   
 \hline 
\end{tabular}
\end{center}
\end{table}

\begin{table}
\begin{center}
\caption{$\sigma$ values for each quadrant in SMICA2 temperature map using symmetric Galactic cuts}
\label{tbl-symmetric} 
\begin{tabular}{ccccc}
\hline\hline
 Symmetric Galactic cut &  $\sigma_{SEQ}$  & $\sigma_{SWQ}$& $\sigma_{NEQ}$ & $\sigma_{NWQ}$  \\
 \hline
  $+/-$ 10$^\circ$                            &2088.76   &529.51    &569.48  &464.12 \\ 
  \hline
  $+/-$ 20$^\circ$                            &1250.25   &635.21    &520.94  &410.29 \\ 
 \hline
  $+/-$ 25$^\circ$                            &730.22    &921.12    &518.85  &333.44 \\ 
 \hline
  $+/-$ 30$^\circ$                            &500.31   &540.26     &289.69   &380.30 \\ 
 \hline 
                   
 \hline 
\end{tabular}
\end{center}
\end{table}

\begin{table}
\begin{center}
\caption{Comparison of the histograms statistics (skewness and kurtosis) in the regions where UT78 is not masked and each of the other masks are masked using SMICA2 and the mean values for the MC simulations}
\label{tbl-kurt} 
\begin{tabular}{crrrrrrrr}
\hline\hline
Mask & SEQ$_{SMICA2}$ & SEQ$_{MC}$ & SWQ$_{SMICA2}$ & SWQ$_{MC}$ & NEQ$_{SMICA2}$& NEQ$_{MC}$ & NWQ$_{SMICA2}$ &NWQ$_{MC}$  \\
 \hline
 UT78- mask-rulerminimal               &(1.53, 2.63) &(1.47, 2.15) &(1.23, 1.24)  &(1.38, 1.89)   &(1.58, 3.14) &(1.41, 1.94)  &(1.64, 3.99) &(1.36, 1.78)  \\

 UT78 - U73                                     &(1.76, 3.83) &(1.52, 2.28) &(1.58, 2.89)  &(1.50, 2.22)   &(1.94, 4.94) &(1.49, 2.21)  &(1.58, 2.77) &(1.50, 2.21)  \\

 UT78 - U66                                   &(1.54, 2.46) &(1.57, 2.46) &(1.47, 1.92)  &(1.56, 2.42)   &(1.76, 3.57) &(1.56, 2.42)  &(1.55, 2.50) &(1.54, 2.37)  \\ 

\hline \hline
\end{tabular}
\end{center}

\end{table}

\begin{table}
\begin{center}
\caption{$\sigma$ values for each circular region in SMICA2 temperature map considering different radius}
\label{tbl-sigma-circ} 
\begin{tabular}{ccccccc}
\hline\hline
         Region & radius & $\sigma$  &radius & $\sigma$  & radius & $\sigma$    \\
  \hline
 1 &$77^{\circ}$ &1474.08 &$80^{\circ}$ &1923.46 &$83^{\circ}$ &1485.98 \\ 
 \hline
 2 &$60^{\circ}$ &328.22 &$70^{\circ}$ &366.67 &$80^{\circ}$ &386.83 \\ 
 \hline
 2 &$60^{\circ}$ &414.50 &$70^{\circ}$ &384.00 &$80^{\circ}$ &524.98 \\ 
 \hline

\end{tabular}
\end{center}
\end{table}

\begin{table}
\begin{center}
\caption{Probabilities of finding in exactly the same circular region of the simulated MC maps bigger or smaller $\sigma$ values than in Planck SMICA2 map using mask-rulerminimal.}
\label{tbl-prob-circ-exac} 
\begin{tabular}{cccr}
\hline\hline
 Region & radius & $\sigma$  & Probalility \\
  \hline
 1  & $80^{\circ}$ &1923.50 & 1.0\% \tablefootmark{a}    \\ 
 \hline
 2  & $60^{\circ}$ &328.22 & 2.1 \% \tablefootmark{b}  \\ 
 \hline
 3  &$70^{\circ}$  &384.00 & 5.3 \%  \tablefootmark{b}    \\ 
 \hline

\end{tabular}
\end{center}
\tablefoottext{a}{Probability of finding $\sigma_{MC}>\sigma_{1}$ }
\tablefoottext{b}{Probability of finding $\sigma_{MC}<\sigma_{2,3}$}
\end{table}

\end{document}